\documentclass[pdftex,journal=jctcce,manuscript=article]{achemso}
\setkeys{acs}{
  keywords      = false
}

\usepackage[T1]{fontenc}
\usepackage[utf8]{inputenc}
\usepackage{float}
\usepackage{color}

\usepackage{amsfonts}
\usepackage{amssymb}
\usepackage{amsmath}
\usepackage{upgreek}
\usepackage{latexsym}
\usepackage{bm}
\usepackage{xspace}

\usepackage{float}
\usepackage{setspace}
\usepackage{graphicx}

\usepackage{lscape}
\usepackage{times}
\usepackage{epsfig}
\usepackage{commath}
\usepackage[normalem]{ulem}
\usepackage{soul}

\usepackage[english]{babel}
\usepackage{wrapfig}

\usepackage[super,sort&compress,numbers]{natbib}

\providecommand{\boldsymbol}[1]{\mbox{\boldmath $#1$}}

\usepackage{microtype}
\usepackage[colorlinks,linkcolor=blue,citecolor=blue,urlcolor=blue]{hyperref}

\title[]{Disentangling enhanced diffusion and ballistic motion of excitons
coupled to Bloch surface waves
with molecular dynamics simulations
}


\author{Ilia Sokolovskii} \email{ilia.sokolovskii@jyu.fi}
\affiliation[Uni Jyvaeskylae, NanoChem]{ Nanoscience Center and
  Department of Chemistry, University of Jyv\"{a}skyl\"{a}, P.O. Box 35, 40014
  Jyv\"{a}skyl\"{a}, Finland.  }

\author{Yunyi Luo}
\affiliation[Uni Jyvaeskylae, NanoChem]{ Nanoscience Center and
  Department of Chemistry, University of Jyv\"{a}skyl\"{a}, P.O. Box 35, 40014
  Jyv\"{a}skyl\"{a}, Finland.  }

\author{Gerrit Groenhof} 
\affiliation[Uni Jyvaeskylae, NanoChem]{ Nanoscience Center and
  Department of Chemistry, University of Jyv\"{a}skyl\"{a}, P.O. Box 35, 40014
  Jyv\"{a}skyl\"{a}, Finland.  }
  
\begin{document}

\setlength{\marginparwidth}{3cm} 

\begin{abstract}
\noindent Placing an organic material on top of a Bragg mirror can significantly enhance exciton transport. Such enhancement has been attributed to strong coupling between the evanescent Bloch surface waves  (BSW) on the mirror, and the excitons in the material. In this regime, the BSW and excitons hybridize into Bloch surface wave polaritons (BSWP), new quasi-particles with both photonic and excitonic character. While recent experiments unveiled a mixed nature of the enhanced transport, the role of the material degrees of freedom in this process remains unclear. To clarify their role, we performed atomistic molecular dynamics simulations of an ensemble of Methylene blue molecules, a prototype organic emitter, strongly coupled to a BSW. In contrast to the established static models of polaritons, even with disorder included, our dynamic simulations reveal a correlation between the photonic content of the BSWP and the nature of the transport. In line with  experiment, we find ballistic motion for polaritons with high photonic character, and enhanced diffusion if the photonic content is low. Our simulations furthermore suggest that the diffusion is due to thermally activated vibrations that drive population transfer between the stationary dark states and mobile bright polaritonic states.
\end{abstract}

\section{Introduction}
The propagation of Frenkel excitons in organic materials is a diffusion process in which the excitons hop between adjacent molecules via dipole-dipole coupling (Förster mechanism)~\cite{Foerster1949} or wave function overlap (Dexter mechanism)~\cite{Dexter1953}. Because the efficiency of these mechanisms depends on the intermolecular separation and orientiation, structural disorder has a negative impact on exciton transfer, resulting in a limited diffusion length of Frenkel excitons, typically below 10~nm\cite{Mikhnenko2015}.

The exciton propagation distance can be increased by placing the organic molecules in a confined electromagnetic field, as found inside an optical microcavity, or near a plasmonic surface\cite{
Sandik2024}. There, excitons can strongly interact with the confined light modes of the optical structure. If the strength of this interaction exceeds the rates associated with losses in the system, the excitons and confined light modes hybridize into polaritons\cite{Torma2015,Rider2022}, which inherit the properties of both constituents of the interaction, including dispersion and hence group velocity. This allows for a long-range propagation of polaritons beyond the diffusion length of Frenkel excitons, as has been demonstrated in various types of optical structures\cite{Coles2014,Lerario2017,Rozenman2018,Georgiou2018,Zakharko2018,Forrest2020,Pandya2021,Pandya2022,Berghuis2022,Xu2022,Balasubrahmaniyam2023}.

Theory has provided important insights into the mechanisms by which strong coupling can enhance exciton transport\cite{Agranovich2007,Litinskaya2008,Michetti2008b,Schachenmayer2015,Feist2015,Allard2022,Ribeiro2022,Ribeiro2023,Osipov2023,Engelhardt2023,Tutunnikov2024,Nitzan2024}, but the description of the material in these works has been limited to two-level systems. To go beyond such simplified model systems and consider the vibrational degrees of freedom, we developed a simulation model based on multi-scale molecular dynamics,\cite{Luk2017,Tichauer2021} in which the structural details of the material are explicitly included. With such simulations, we could demonstrate that polariton wave packets propagate in a diffusive manner due to reversible population exchanges between stationary dark states and propagating bright states, which get populated along the whole lower polariton branch\cite{Sokolovskii2023,Sokolovskii2024b}. While experiment\cite{Shi2014} and theory\cite{Michetti2005,Engelhardt2023,Chng2024} suggest that the coherence of polaritons, and hence the propagation distance, increases with their photonic weight, it remains unclear whether propagation of individual wave packets constituting the total polaritonic wave packet, is fully ballistic or might also be diffusive depending on the wave vector at which an individual wave packet is formed.

Recently, this question was addressed in two separate experimental studies~\cite{Balasubrahmaniyam2023,Xu2022}. In the first study, conducted by Balasubrahmaniyam \textit{et al.}, spatiotemporal ultrafast pump-probe microscopy was used to track polariton transport in a system of 5,5',6,6'-tetrachloro-1,1'-diethyl-3,3'-di(4-sulfobutyl)-benzimidazolocarbocyanine (TDBC) J-aggregates deposited on top of a distributed Bragg reflector (DBR) supporting Bloch surface waves (BSWs)~\cite{Balasubrahmaniyam2023}. By probing the differential reflectivity, $\Delta R/R$, at different angles and energies as a function of time after off-resonant excitation into the J-aggregates, a transition from ballistic propagation with a velocity close to the corresponding polariton group velocity, to diffusive propagation with a much lower velocity, was observed when the photonic contribution to the polaritonic states decreased. This transition was attributed to a competition between molecular-scale disorder and long-range correlation due to strong coupling, with the former prevailing at small photonic fractions.

In the second study, carried out by Xu \textit{et al.}\cite{Xu2022}, polariton transport was observed in a variety of inorganic exciton-cavity structures by means of a momentum-resolved ultrafast polariton imaging technique. As in Balasubrahmaniyam \textit{et al.}\cite{Balasubrahmaniyam2023}, both a deviation of the propagation speed from the polariton group velocity and a transition from ballistic to diffusive transport was observed as the photonic contribution to the polaritonic states decreased and was attributed to a more intensive scattering by lattice phonons. 


Thus, the results of the two experiments agree that polariton transport undergoes a crossover between ballistic and diffusion regimes when the polariton states become more exciton-like. However, in the case of organic molecules, it remains unclear whether this crossover is due to reversible non-adiabatic population transfers between bright and dark states, or whether such a transition can be caused solely by the structural molecular disorder. To address this question, 
we mimic the experiment of Balasubrahmaniyam \textit{et al.}\cite{Balasubrahmaniyam2023} by means of multiscale quantum mechanics/molecular mechanics (QM/MM) molecular dynamics (MD) simulations~\cite{Warshel1976b,Luk2017}. The results of our simulations suggest that the change in the transport regime is, indeed, caused by vibrationally induced non-adiabatic population exchanges between polaritonic and dark states, which underscores the importance of molecular vibrations in the transport of organic exciton-polaritons.

\section{Theoretical background}

\begin{figure}[!thb]
\centering
\includegraphics[width=0.7\textwidth]{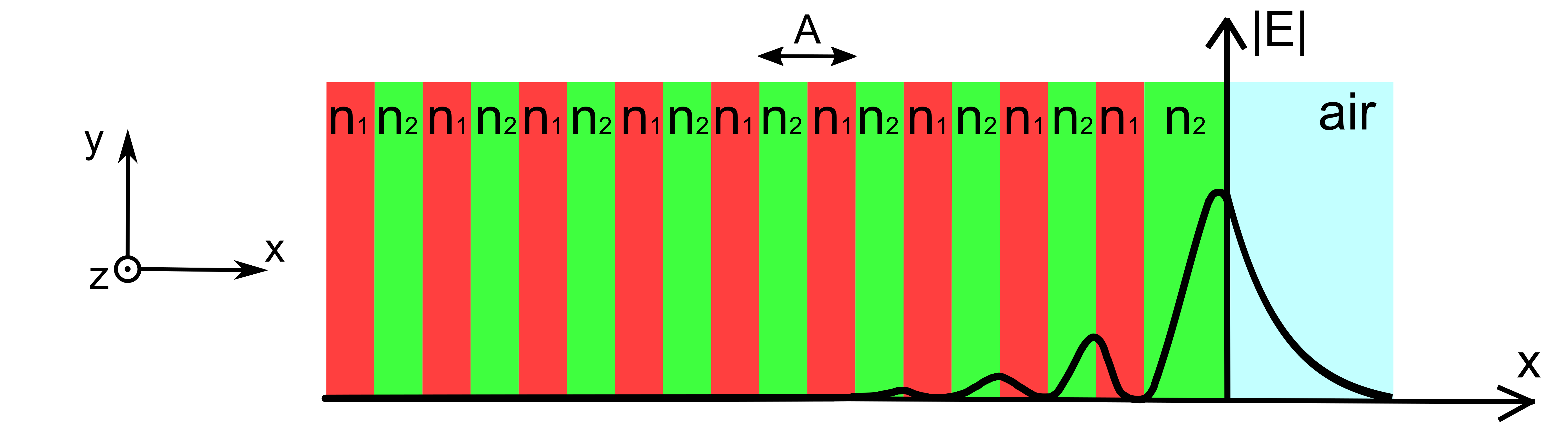}
  \caption{Schematic representation of a distributed Bragg reflector (DBR) supporting a Bloch surface wave (BSW). The DBR consists of alternating layers with different refractive indices, $n_1$ and $n_2$, with period $A$ along the $x$-axis. Additionally, a surface defect layer with refractive index $n_2$ and a thickness different from the thicknesses of the other layers is introduced. The electric field strength distribution of the BSW is shown as a black line.}
  \label{fig:BSW_scheme}
\end{figure}

Before discussing the details of our MD simulations, we briefly discuss the origin of the formation of BSWs in a distributed Bragg reflector (DBR). As illustrated in Figure~\ref{fig:BSW_scheme}, a DBR is a one-dimensional photonic crystal, a structure in which the dielectric constant varies periodically, \textit{i.e.}, $\varepsilon(x)=\varepsilon(x+A)$ with period $A$. Just as the periodicity of ions in an atomic crystal leads to the appearance of allowed bands and band gaps, the periodicity of the dielectric constant in the DBR leads to the appearance of so-called pass-bands and stop-bands. For an electro-magnetic (EM) wave with a frequency inside a stop-band, multiple reflections of this wave from the boundaries between layers with different dielectric constants, result in the occurrence of destructive interference, which makes it impossible for the light to travel through the DBR. However, the introduction of a defect, such as a layer with a thickness or dielectric constant different from that of the other layers, may allow for localised states to appear in the stop-band~\cite{Sakoda2005}. 

In addition to EM waves propagating within the volume, DBRs support (i) surface waves that propagate in both dielectric and air (states in pass-bands above the light line); (ii) surface waves that propagate in the dielectric while decaying in air (states in pass-bands below the light line), and (iii) states that  decay within the dielectric while propagating in air (states in stop-bands above the light line)~\cite{Meade1991}. The introduction of a surface defect results in the appearance of a fourth type of surface wave, which is a wave localised in both dielectric and air~\cite{Robertson1993}. This is the Bloch surface wave. 

Because BSWs exist within the stop-bands and below the light line~\cite{Meade1991}, \textit{i.e.}, at angles larger than the critical angle for total internal reflection, the radiation cannot be emitted into free space, which results in an extremely long lifetime that is significantly higher than for 
typical Fabry-P\'{e}rot microcavities with metallic mirrors, or plasmonic structures. As a consequence, the lifetime of polaritons formed due to strong coupling between excitons and BSWs, can reach several hundreds to thousands of femtoseconds~\cite{Lerario2017,Forrest2020,Rashidi2024}. 

Because the BSW is a surface wave, evanescent in the direction perpendicular to the surface (\textit{i.e.}, $x$ in Figure~\ref{fig:BSW_scheme}), there is no restriction on the in-plane propagation over the surface of the Bragg mirror 
(\textit{i.e.}, in the $y$- and $z$-directions). Therefore, the electric field distribution of the BSW in the air is defined as  
\begin{equation}
 E(x,y,z)=E_0e^{-|K|x}e^{i(k_yy+k_zz)}
   \label{eq:BSW_field}
\end{equation}
with $E_0$ the amplitude of the electric field at the surface of the DBR (\textit{i.e.}, at $x=0$), and $K$ the complex Bloch wave number~\cite{Koju2017}. 

A major advantage of BSW-polaritons over polaritons in Fabry-P\'{e}rot cavities is the much higher group velocity of the LP branch.
In the BSW structures, the dispersion is close to the light line in free space, and the group velocity of the LP branch can approach the speed of light. 
This results in a tremendous propagation of BSW-polaritons reaching tens to hundreds micrometers~\cite{Lerario2017,Barachati2018,Forrest2020,Balasubrahmaniyam2023}.

\section{Simulation details}

In this work, we perform QM/MM molecular dynamics simulations of $N=1024$ Methylene blue molecules (MeB, Figure~\ref{fig:figure_2}\textbf{a}) in water coupled to a BSW. We note that in our simulations, MeB molecules were used instead of the J-aggregates that were used in the experiment~\cite{Balasubrahmaniyam2023}, as the latter's complexity makes their simulation in a cavity currently intractable. Nevertheless, in spite of their larger absorption line-width, the MeB molecules share the most important features of the J-aggregates, namely a bright electronic transition and several vibrational modes that are Raman-active~\cite{Somaschi2011}. The electronic ground (S$_0$) and  first excited state (S$_1$) of MeB are modeled with density functional theory (DFT)~\cite{Hohenberg1964} and time-dependent DFT (TDDFT)~\cite{Runge1984, Casida1998}, respectively, using the B97 functional~\cite{Becke97} in combination with the 3-21G basis set. The water molecules are described with the TIP3P model~\cite{Jorgensen1983}. At the level of TDDFT theory employed in this work, the excitation energies of MeB in water (2.5 eV) are significantly overestimated with respect to experiment (1.9 eV). Such  discrepancies have also been noted in previous works\cite{Queiroz2021,Dunnett2021}, and were attributed to the large difference in the charge density between the ground and excited state, which is notoriously difficult to describe accurately with TDDFT\cite{LeGuennic2015}. Because in our simulations, we only consider a single layer of molecules, we neglect the $x$-dependence in Equation~\ref{eq:BSW_field}. Furthermore, we also restrict ourselves to model one-dimensional transport along a chain of molecules in the $z$-direction. With these simplifications, the field distribution becomes $E(z)=E_0e^{ik_zz}$.

The 1D BSW is discretized into $n_{\text{modes}}=120$ modes. To obtain a similar polariton dispersion as in the experiment, we fit the experimental dispersion of the BSW with a linear function (dashed-dotted line in Figure~\ref{fig:figure_2}\textbf{b}) and tune it to be resonant with the excitation energy of MeB molecules (\textit{i.e.}, $\hbar\omega_{\text{MeB}}=2.50$~eV at the TDDFT/B97//3-21G level of theory; dashed line in Figure~\ref{fig:figure_2}\textbf{b}) at wave vector $k_z=11.41~\mu$m$^{-1}$. With an electric field strength of $0.071$~MV~cm$^{-1}$, the Rabi splitting between the upper (UP) and lower polariton (LP) branches is $131$~meV, which is close to the Rabi splitting in the experimental study\cite{Balasubrahmaniyam2023}. 

In our MD simulations, we apply the Born-Oppenheimer approximation~\cite{Galego2015,Luk2017} to separate the nuclear degrees of freedom, which are treated classically, from the electronic plus photonic degrees of freedom, which are treated quantum mechanically with the QM/MM extension~\cite{Tichauer2021} of the Tavis-Cummings Hamiltonian\cite{Jaynes1963,Tavis1969}:
\begin{equation}
\begin{array}{ccl}
    {\hat{H}}^{\text{TC}} &=& \sum_j^{N} \hbar\omega_{\text{exc}}({\bf{R}}_j)\hat{\sigma}^+_j\hat{\sigma}^-_j+\sum_{j}^N V_{\text{S}_0}({\bf{R}}_j)+\sum_{p}^{n_\text{modes}}\hbar\omega_{\text{cav}}({\bf{k}}_{z,p})\hat{a}_p^\dagger\hat{a}_p\\
    \\
    &&-\sum_j^N\sum_p^{n_\text{modes}}\sqrt{\frac{\hbar\omega_{\text{cav}}(\textbf{k}_{z,p})}{2\epsilon_0V}}
    {{\boldsymbol{\mu}}}({\bf{R}}_j)\cdot\left[e^{ik_{z,p} z_j}\hat{\sigma}_j^+\hat{a}_p+e^{-ik_{z,p} z_j}\hat{\sigma}_j^-\hat{a}_p^\dagger\right],
\end{array}
\label{eq:TC_Ham}
\end{equation}
where $\hat{\sigma}^+_{j} = |\text{S}_1^j\rangle\langle\text{S}_0^j|$ is an operator that excites molecule $j$ with nuclear coordinates ${\bf{R}}_j$ from the electronic ground state $|\text{S}_0^j\rangle$ with energy $V_{\text{S}_0}({\bf{R}}_j)$ into the first electronic excited state $|\text{S}_1^j\rangle$ with energy $V_{\text{S}_1}({\bf{R}}_j)$. Accordingly, the excitation energy is defined as $\hbar\omega_{\text{exc}}({\bf{R}}_j)=V_{\text{S}_1}({\bf{R}}_j)-V_{\text{S}_0}({\bf{R}}_j)$. Likewise, $\hat{\sigma}^-_{j} =|\text{S}_0^j\rangle\langle\text{S}_1^j|$ de-excites molecule $j$ from electronic excited state $|\text{S}_1^j\rangle$ into the electronic ground state $|\text{S}_0^j\rangle$. Operators $\hat{a}^\dagger_p$ and $\hat{a}_p$ create and annihilate a photon of energy $\hbar\omega_{\text{cav}}({\bf{k}}_{z,p})$ in BSW mode $p$ with in-plane momentum ${\bf{k}}_{z,p}$ along the surface of the DBR. Finally, $
{{\boldsymbol{\mu}}}({\bf{R}}_j)$ is the transition dipole moment of molecule $j$, and $z_j$ is the position of molecule $j$ on the surface of the DBR structure. 

The Ehrenfest molecular dynamics approach is used to model the classical degrees of freedom~\cite{Ehrenfest1927}, which evolve on a potential energy surface that is the expectation value of the energy of the total wave function of the quantum degrees of freedom: $V({\bf{R}})=\langle\Psi|\hat{H}|\Psi\rangle$. 
The total wave function, $|\Psi(t)\rangle$, is propagated along the classical trajectory as a linear combination of diabatic product states between the $N$ molecular excitations and $n_{\text{modes}}$ BSW modes~\cite{Sokolovskii2024b}:
\begin{equation}
\vert\Psi(t)\rangle=\sum_j^{N+n_\text{modes}} d_j(t)\vert\phi_j\rangle\label{eq:totalwf},
\end{equation}
with 
\begin{equation}
|\phi_j\rangle = \hat{\sigma}_j^+|\text{S}_0^1\text{S}_0^2..\text{S}_0^{N-1}\text{S}_0^N\rangle\otimes|00..0\rangle\label{eq:basis1}
\end{equation}
for $1\le j\le N$, and 
\begin{equation}
|\phi_{j >N}\rangle = \hat{a}_{j-N}^\dagger|\text{S}_0^1\text{S}_0^2..\text{S}_0^{N-1}\text{S}_0^N\rangle\otimes|00..0\rangle\label{eq:basis2}
\end{equation}
for $N < j\le N+n_\text{modes}$. State $|\phi_j\rangle$ corresponds to molecule $j$ in its first electronic excited state (S$^j_1$), while the other molecules are in the ground state (S$_0^{i\neq j}$) and the photonic states are empty, whereas state $|\phi_{j>N}\rangle$ corresponds to one of the BSW modes excited, with all molecules in the electronic ground state. In Equation~\ref{eq:totalwf}, the $d_j(t)$ are the time-dependent expansion coefficients of the total wave function, which reflect the population of each molecular excitation and each photonic mode during the evolution of the system.  These coefficients are propagated with a unitary propagator~\cite{Granucci2001}. Further details of the simulation method can be found in the SI (Section 1) or in previous publications~\cite{Luk2017,Tichauer2021,Sokolovskii2024b}.


To mimic the initial conditions of the experiment, in which a single J-aggregate was optically pumped~\cite{Balasubrahmaniyam2023}, we prepare the MeB-BSW system in the first excited electronic  state (S$_1$) of a single MeB molecule, $j$, located at $z_j=125~\mu$m (\textit{i.e.}, $d_j(0)=1$ and $d_{i\neq j}(0)=0$ in Equation~\ref{eq:totalwf}), which is at the centre of a periodic DBR surface of width $L_z=250~\mu$m (Figure~\ref{fig:figure_2}{\textbf{a}). A total of five Ehrenfest QM/MM trajectories are computed for 200~fs with an integration timestep of 0.5~fs. Because these MD trajectories are run for much shorter than the typical lifetime of a BSW, we neglect decay in our simulations. The temperature was kept constant at 300~K with the v-rescale thermostat~\cite{Bussi2007}.  Further details of the simulations performed in this article are presented in the Supporting Information (SI).


For the analysis of the trajectories, we also
expand the total time-dependent wave function in the basis of the eigenstates of the Tavis-Cummings Hamiltonian, as follows:
\begin{equation}
\vert\Psi(t)\rangle=\sum_j c_m(t)\vert\psi_m\rangle\label{eq:totalwf_adia},
\end{equation}
where
\begin{equation}
\vert\psi_m\rangle=\left(\sum_{j}^N\beta^m_j\hat{\sigma}^+_j + \sum_{p}^{n_{\text{modes}}}\alpha^m_p\hat{a}_p^\dagger\right)\vert\text{S}_0^1\text{S}_0^2..\text{S}_0^{N-1}\text{S}_0^{N}\rangle\otimes\vert0\rangle.\label{eq:Npolariton}
\end{equation}
Here, the $\beta_j^m$ and $\alpha_p^m$ expansion coefficients denote contributions of the molecular excitons ($\vert\text{S}_1^j\rangle$) and of the photonic modes ($\vert 1_p\rangle$) to adiabatic eigenstate $\vert\psi_m\rangle$. These coefficients are obtained by diagonalizing the matrix representation of $\hat{H}^\text{TC}$ (Equation~\ref{eq:TC_Ham}) in the basis of the diabatic product states (Equation~\ref{eq:basis1}-\ref{eq:basis2}). The time-dependent expansion coefficients, $c_m(t)$, in this adiabatic representation are thus related to the time-dependent expansion coefficients, $d_j(t)$, in the diabatic representation (Equation~\ref{eq:totalwf}) via the unitary matrix, ${\bf{U}}$, that diagonalizes the Tavis-Cummings matrix (\textit{i.e.}, $c_m(t) = \sum_j^{N+n_\text{modes}} U^\dagger_{mj}d_j(t) $, with $U_{jm} = \beta_j^m$ if $j\le N$ and $U_{jm} = \alpha_{j-N}^m$ if $j>N$).


To explore how the photonic character of the adiabatic eigenstates affects their contribution to the overall transport, we decompose the total wave function into \textit{partial} wave functions, $|\Psi^{\text{part}}_{\text{phot},w}(t)\rangle$, within a narrow range of wave vectors. These partial wave functions are linear combinations of the eigenstates 
(Equation~\ref{eq:Npolariton}) in fixed intervals of wave vectors (or equivalently, energies). Each interval, called a window, $w_i$, is centered at $k_{z,i}$ and ranges from $k_{z,i}^\text{min}$ to $k_{z,i}^\text{max}$. The expansion coefficients, $c_{m\in w_i}(t)$, of these partial wave functions are thus obtained by projecting the adiabatic states within a window onto the total time-dependent wave function in the adiabatic representation (Equation~\ref{eq:totalwf_adia}).

In this analysis, we consider only the elements associated with the BSW modes, \textit{i.e.}, the second term in Equation~\ref{eq:Npolariton}. 
This choice was made because 
the excitonic part of the total wave function is more responsive to the excitation energy disorder\cite{Agranovich2007}, and unambiguous resolution of the transport regime based on the full partial wave function, which includes both excitonic and cavity components (\textit{i.e.}, both terms in Equation~\ref{eq:Npolariton}), would have required averaging over many more simulations than we can currently afford to run (five). Nevertheless, as we show in detail in the SI (Section~3.1), this choice does not change the conclusions that we draw below, as the analysis of the propagation of a partial wave function that includes both excitonic and photonic elements, yields very similar results, albeit with higher noise levels.


\section{Results and Discussion}

In Figure~\ref{fig:figure_2}\textbf{c}, we show the time evolution of the probability density of the total polaritonic wave function, $|\Psi(t)|^2$, after excitation of a single MeB molecule, located at 125~$\mu$m on the DBR surface. The wave packet rapidly expands, with the front of the wave packet moving at the group velocity of the lower polariton, $\upsilon_{\text{LP}}^{\text{max}}=173~\mu$m~ps$^{-1}$ (dashed line in Figure~\ref{fig:figure_2}\textbf{c} and Figure~S2, SI), and the tail remaining at the position where the molecule was excited. Such spreading of the wave packet suggests that the overall transport mechanism is a combination of ballistic motion and diffusion. To disentangle these two processes, we analyse how polaritonic states contribute to the transport as a function of their photonic content. To perform this analysis, we monitor the propagation of adiabatic states within finite windows of $k_z$-vectors, two of which are illustrated as rectangles in Figure~\ref{fig:figure_2}\textbf{b}.


\begin{figure}[!htb]
\centering
\includegraphics[width=1\textwidth]{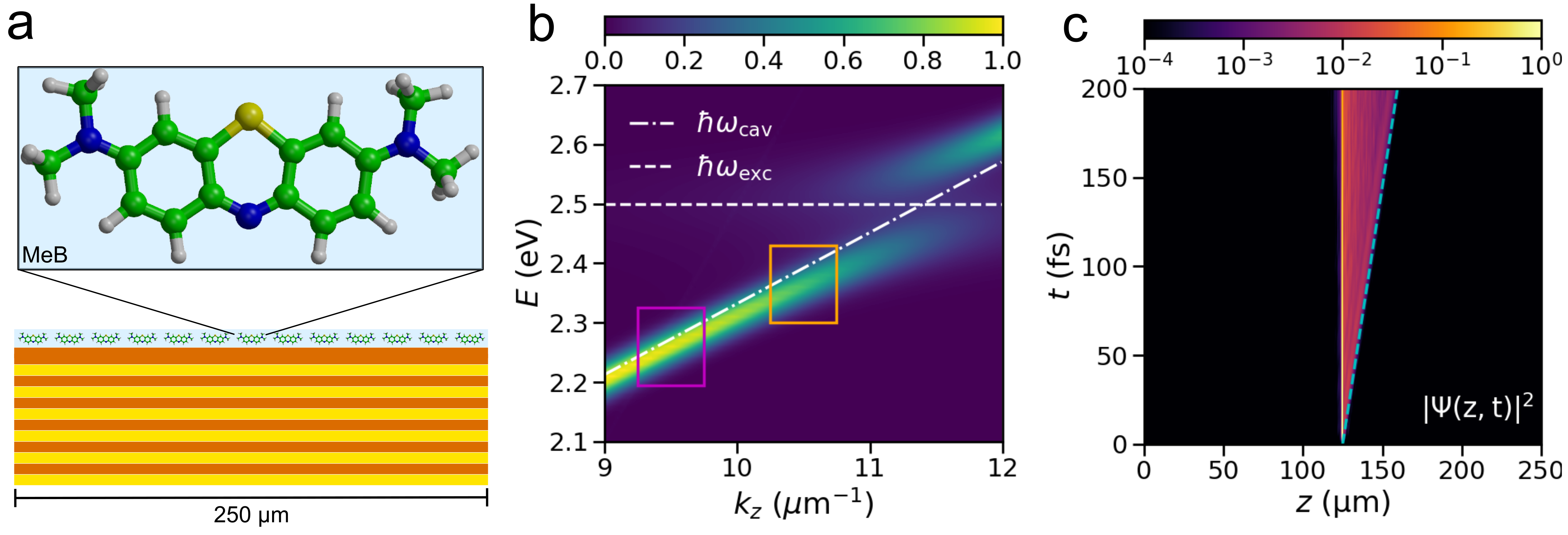}
  \caption{Panel~\textbf{a}: Schematic illustration of MeB molecules deposited on top of a DBR. The electronic ground (S$_0$) and excited (S$_1$) states are calculated at the QM level with density functional theory (DFT) and time-dependent (TD-)DFT, respectively, using the B97 functional and the 3-21G basis set. The solvent molecules (water, not shown) are described at the MM level with the TIP3P model. Carbon atoms are shown in green, nitrogen atoms in blue, sulfur atom in yellow, and hydrogen atoms in grey. Panel~\textbf{b}: Normalised angle-resolved absorption spectrum of the MeB-Bloch surface wave (BSW) system. The dashed line corresponds to the excitation energy of MeB at 2.50~meV at the TDDFT/B97//3-21G level of theory, and the dashed-dotted line shows the BSW dispersion. The purple and orange rectangles illustrate two wave vector windows, from which the partial photonic wave functions, $|\Psi^{\text{part}}_{\text{phot}}\rangle$, are extracted and plotted in Figure~\ref{fig:BSW_transport}. Panel~\textbf{c}: Space-time map of the probability amplitude of the total wave function, $|\Psi(z,t)|^2$. The dashed line indicates the maximum group velocity of the LP branch, $\upsilon_{\text{LP}}^{\text{max}}=173~\mu$m~ps$^{-1}$.}
  \label{fig:figure_2}
\end{figure}

In Figure~\ref{fig:BSW_transport}\textbf{a} and \textbf{b}, we show the propagation of the partial photonic wave functions, $|\Psi_{\text{phot}}^{\text{part}}|^2$, associated with states in two windows: $w_a$ from $k_z =  9.25~\mu$m$^{-1}$ to $9.75~\mu$m$^{-1}$, and $w_b$ from $k_z=10.25~\mu$m$^{-1}$ to $k_z=10.75~\mu$m$^{-1}$ (purple and orange rectangles in Figure~\ref{fig:figure_2}\textbf{b}, respectively). The mean squared displacements (MSD) of these partial photonic wave packets are plotted in Figure~\ref{fig:BSW_transport}\textbf{c}. 
The MSD plots for the other windows, as well as for the total probability density ($|\Psi(z,t)|^2$), are shown in the SI (Section~3.1). By fitting the mean squared displacement to the general expression for MSD,
\begin{equation}
    \text{MSD}(t)=2D_{\beta}t^{\beta},
\end{equation}
we determine whether the propagation is ballistic or diffusive from the value of the transport exponent, $\beta$. Purely ballistic transport corresponds to $\beta=2$, whereas pure diffusion corresponds to the transport exponent equal to unity with $D_{\beta}$ becoming the diffusion coefficient.

\begin{figure}[!htb]
\centering
\includegraphics[width=1\textwidth]{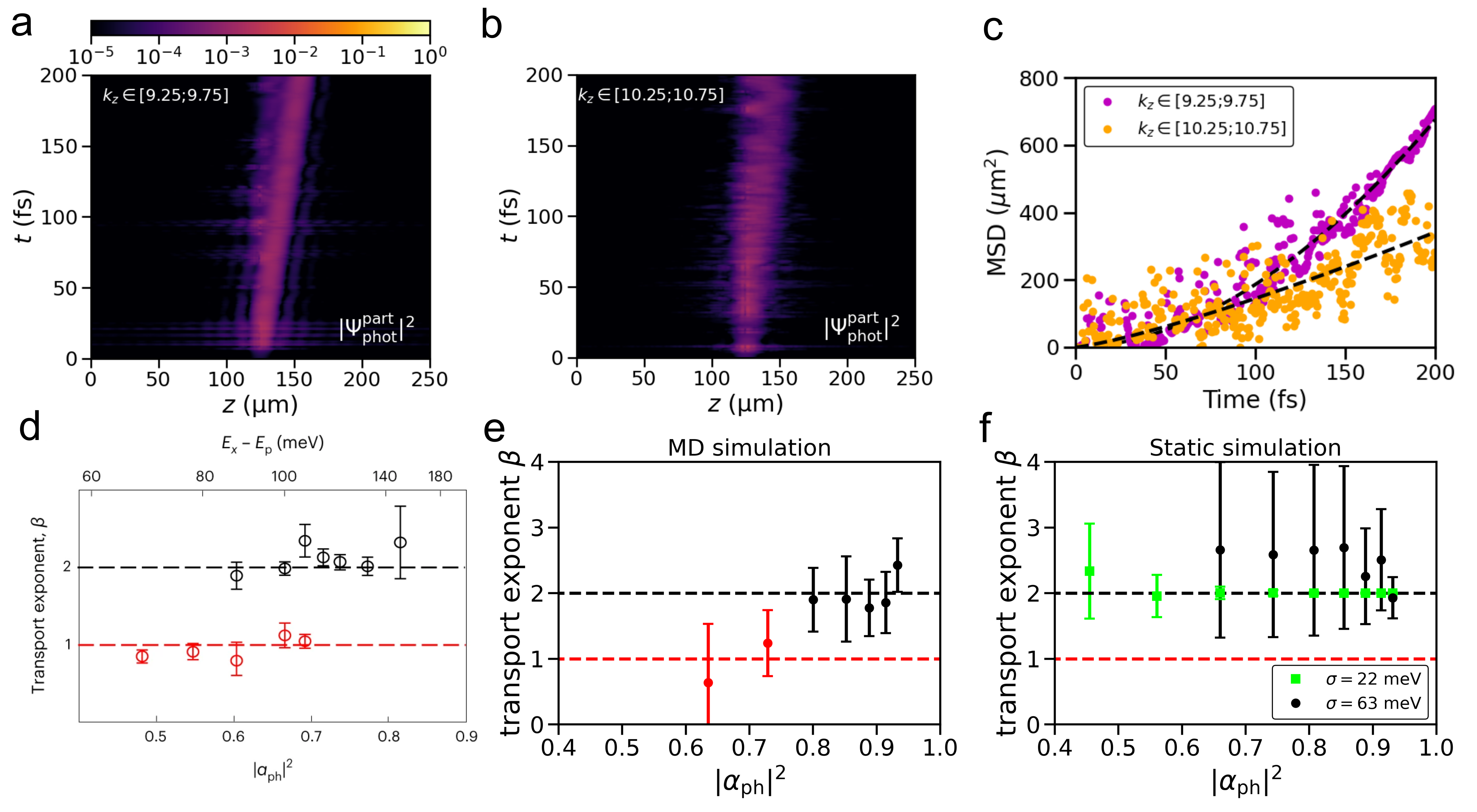}
  \caption{Panels~\textbf{a}-\textbf{b}: Probability density, $|\Psi^{\text{part}}_{\text{phot}}|^2$, of the partial photonic wave function extracted from the two windows depicted as purple (\textbf{a}) and orange (\textbf{b}) rectangles in Figure~\ref{fig:figure_2}\textbf{b}. Panel~\textbf{c}: MSD of the photonic wave function in panels~\textbf{a}-\textbf{b}. The dashed lines correspond to the fit to $\text{MSD}=2D_{\beta}t^{\beta}$ with $\beta$ the transport exponent. Panels~\textbf{d}~and~\textbf{e}: The transport exponent as a function of the cavity modes contribution $|\alpha_{\text{ph}}|^2$ to polaritonic states, extracted from the experiment (\textbf{d}) and MD simulations (\textbf{e}). Panel~\textbf{d} is reproduced with permission from Balasubrahmaniyam \textit{et al.}, \textit{Nat. Mater.}, \textbf{22}, 338–344 (2023). Copyright 2023 Springer Nature Ltd. Panel~\textbf{f}: Values extracted from simulations of two-level systems with 
  static excitation energy disorder of $\sigma=22$~meV (green squares) and $\sigma=63$~meV (black circles). In panels \textbf{e} and \textbf{f}, the transport exponents were extracted from windows of width $\Delta k_z=0.5~\mu$m$^{-1}$ centred between $k_z=9.25~\mu$m$^{-1}$ and $k_z=10.75~\mu$m$^{-1}$ with a step of $0.25~\mu$m$^{-1}$ (Table~S2). The error bars in panels \textbf{e} and \textbf{f} depict the standard deviations of, respectively, five and five hundred individual simulations.}
  \label{fig:BSW_transport}
\end{figure}

In Figure~\ref{fig:BSW_transport}\textbf{e} we plot the value of the transport exponent for the partial photonic wave functions extracted from all windows between $k_z^c=9.25~\mu$m$^{-1}$ and $k_z^c=10.75~\mu$m$^{-1}$, 
as a function of the photonic Hopfield coefficient ($|\alpha_{\text{ph}}|^2$, Equation~\ref{eq:Npolariton}) at the center of the windows (Table~S2).
The plot suggests that shifting the window upwards along the lower polariton branch towards states with a lower photonic content, is accompanied by a change in the transport exponent, $\beta$, from two to one, indicating a transition between ballistic propagation and diffusion, in line with the experiment (Figure~\ref{fig:BSW_transport}\textbf{d})~\cite{Balasubrahmaniyam2023}.

Such transition from ballistic to diffusive transport cannot be reproduced in simulations of two-level systems with a static
excitation energy disorder. In these simulations (Section~3.2 in the SI), the 
disorder was modelled by randomly drawing the excitation energies of the two-level systems from a Gaussian distribution 
\begin{equation}
  p(E)=\frac{1}{\sqrt{2\pi}\sigma}\exp\left[-\frac{\left(E-E_0\right)^2}{2\sigma^2}\right],
\label{eq:guass_distr}
\end{equation} 
where $E_0$ is the mean value, 
and $\sigma$ is the "disorder strength" that  determines the absorption line-width of the disordered two-level system.
As shown in Figure~\ref{fig:BSW_transport}\textbf{f}, the transport exponent remains close to $\beta=2$ for the full range of LP states with a well-defined wave vector\cite{Agranovich2003,Litinskaya2004}. For $\sigma=63$~meV, which matches the line-width of the MeB QM/MM model used in our atomistic MD simulations, LP states are well defined for states with $|\alpha_{\text{ph}}|^2>0.56$, whereas for $\sigma=22$~meV, which is typical of J-aggregates~\cite{Walker2002,Bricks2007}, the LP is well-defined for states with  $|\alpha_{\text{ph}}|^2>0.36$. 

The difference between the transport mechanisms
observed in the MD and 
static simulations (Figure~\ref{fig:BSW_transport}\textbf{e} and \textbf{f}),  suggests a decisive role of molecular vibrations in modifying the mechanism of polariton transport along the lower polariton branch. Through displacements along the non-adiabatic coupling vector, these vibrations drive population transfer between polaritonic states~\cite{Tichauer2022}. Because the strength of the non-adiabatic coupling is inversely proportional to the energy gap between these states~\cite{Yarkony2012}, population transfer between the stationary dark states, which are distributed around the molecular absorption maximum, on the one hand, and the low-energy, highly-photonic LP states on the other hand, is unlikely, resulting in a scattering-free ballistic propagation of population in these states. Moving up in energy along the LP branch towards higher $k_z$-vectors, reduces the energy gap.
Therefore, the transfer rate, and hence the likelihood of population getting transiently trapped in the dark state manifold, increases  significantly. Eventually, as the energy gap narrows further, and the  polaritonic states start overlapping with the dark states in the exciton reservoir~\cite{Groenhof2019,Dutta2023,schwennicke2024}, the population exchange between the propagating bright and stationary dark states  becomes continuous and reversible, rendering the 
propagation less ballistic and more diffusive.

To further elucidate the role of 
molecular vibrations for the observed crossover between regimes of polariton transport, we repeated the MD simulations of $N=1024$ MeB molecules in vacuum, with constraints imposed on all bond lengths~\cite{Hess1997} and on the out-of-plane motions of the heavy atoms (Section 3.3 in the SI).
By restricting the displacements along the vibrational modes, these constraints reduce the absorption line-width and hence the overlap between the dark states and LP branch. In addition, because the non-adiabatic coupling between the bright polaritonic states ($\psi_m$) and dark states ($\psi_l$) depends on the velocity of displacements that overlap with the non-adiabatic coupling vector (\textit{i.e.}, $V_{m,l}={\bf{d}}_{m,l}\cdot\dot{\bf{{R}}}$, with $\textbf{d}_{m,l} = \frac{\langle \psi_l|\nabla_{\textbf{R}}{\hat{H}}^{\text{TC}}|\psi_m\rangle}{E_m-E_l}$ and ${\dot{\bf{R}}}$ the velocity)\cite{Tichauer2022}, the constraints also reduce the non-adiabatic coupling strength and, therefore, the rates at which population transfers. As a consequence, population transfer into the dark states is suppressed, the transport becomes ballistic over the full range of the LP branch, and no  crossover into a diffusion regime is observed (Figure~S10).

In addition, we also performed simulations of two-level systems with quasi-dynamic disorder\cite{Nitzan2023}, in which excitation energies were drawn from the Gaussian distribution (Equation~\ref{eq:guass_distr}) and resampled every five femtoseconds  in order to imitate the continuous energy re-distribution due to dynamics of molecules  (Section~S3.4 in the SI). Unlike the molecular dynamics simulations, no clear transition between the transport regimes was observed in these quasi-dynamic simulations, and the transport exponent, $\beta$, remained closer to two for most of the $k_z$ -vectors windows (Figure~S11). This finding indicates that the transition cannot be caused by the excitation energy 
disorder only and hence further emphasizes the role of molecular vibrations in this process.


Finally, to confirm that dark states are essential for the crossover between ballistic transport and diffusion, we also ran MD simulations of a system with a minimal number of dark states. This system was created by including as many molecules as photonic modes, \textit{i.e.}, $N=120$ MeB molecules and $n_{\text{modes}}=120$ BSW modes. Without disorder, there are no dark states, and all states are polaritonic (Section~3.5 in the SI). In these simulations, the populations of excitons and BSW modes become approximately equal (Figure~S12), indicating that most population resides in bright polaritonic states, and that the population transfer from these bright states back into the dark state manifold is minimal, in line with previous work that demonstrated a dependence of the population transfer rates on the density of states\cite{delPino2015,Ulusoy2019,Eizner2019,Davidsson2023,Borjesson2023,Perez2024,Sokolovskii2024c}. Without significant population transfer into stationary dark states, there is no clear transition from ballistic transport to diffussion within the range of polaritonic states with a well-defined $k_z$ vector (Figure~S13). Thus, to capture the crossover that was observed experimentally for BSW-polaritons\cite{Balasubrahmaniyam2023}, the ratio between dark and bright polaritonic states needs to be sufficiently large (\textit{i.e.}, $N_{\text{dark}}/N_{\text{bright}}\gg1$) as is the case in most experiments.



\section{Conclusion}
To summarize, we have performed molecular dynamics simulations of polariton transport in a system of Methylene blue molecules strongly coupled to the Bloch surface wave on a distributed Bragg reflector. In line with experiment\cite{Balasubrahmaniyam2023}, our simulations reveal a transition between ballistic and diffusive propagation as the photonic contribution to the polaritonic states decreases. Importantly, no transition was observed in simulations, in which the molecules were frozen and modeled as two-level systems. The absence of a transition in such model systems, underscores the key role of molecular vibrations in changing the transport regime along the LP branch.


\section* {Author Information}
\subsection*{Corresponding Author}
Email: ilia.sokolovskii@jyu.fi

\subsection*{ORCID}

\noindent I. Sokolovskii: 0000-0003-3367-0660

\noindent G. Groenhof: 0000-0001-8148-5334

\subsection*{Funding}
This work was supported by the Academy of Finland (Grants 332743 and 363659).

\subsection*{Notes}
The authors declare no competing financial interests.

\section*{Acknowledgements}
We thank Tal Schwartz and M. Balasubrahmaniyam for fruitful discussions and for sharing the BSW dispersion data. We thank the Center for Scientific Computing (CSC-IT Center for Science) for very generous computer resources.

\bibliography{main}

\end{document}


\setlength{\marginparwidth}{3cm} 
\newpage
\tableofcontents

\newpage

\section{Molecular Dynamics Simulation Model}

\subsection{Multiscale Tavis-Cummings Hamiltonian}

In our model, strong coupling between \textit{electronic} excitations of molecules and light modes of a cavity is described within the Born-Oppenheimer approximation, in which fast electronic \textit{plus} photonic degrees of freedom are separated from slow nuclear degrees of freedom~\cite{Galego2015}. To describe the electronic-photonic degrees of freedom, we extend the traditional Tavis-Cummings Hamiltonian~\cite{Jaynes1963,Tavis1969} to the case of multiple cavity modes, $n_{\text{modes}}$, and replace two-level systems with adiabatic electronic states of the molecules~\cite{Tichauer2021}:
\begin{equation}
\begin{array}{ccl}
    {\hat{H}}^{\text{TC}} &=& \sum_j^N \hbar\omega_{\text{exc}}(\mathbf R_j)\hat{\sigma}^+_j\hat{\sigma}^-_j+\sum_{k_z}^{n_\text{modes}}\hslash\omega_{\text{cav}}(k_z)\hat{a}_{k_z}^\dagger\hat{a}_{k_z}+\\
    \\
    &&\sum_j^N\sum_{k_z}^{n_\text{modes}}\hslash g_j(k_z) \left(\hat{\sigma}^+_j\hat{a}_{k_z}e^{ik_zz_j}+\hat{\sigma}^-_j\hat{a}_{k_z}^\dagger e^{-ik_zz_j}\right)+\\
    \\
    &&\sum_{i}^N V^\text{mol}_{\text{S}_0}({\bf{R}}_i).
    \end{array}
    \label{eq:dTCH}
\end{equation}

In Equation~\ref{eq:dTCH}, the operator $\hat{\sigma}_j^+$  ($\hat{\sigma}_j^-$) excites (de-excites) molecule $j$ from the electronic ground (excited) state $|\text{S}_0^j({\bf{R}}_j)\rangle$ ($|\text{S}_1^{j}({\bf{R}}_j)\rangle$) to the electronic excited (ground) state $|\text{S}_1^{j}({\bf{R}}_j)\rangle$ ($|\text{S}_0^j({\bf{R}}_j)\rangle$); ${\bf{R}}_j$ is the vector of the Cartesian coordinates of all atoms in molecule $j$, centered at $z_j$;
the operators $\hat{a}_{k_z}^\dagger$ and $\hat{a}_{k_z}$, respectively, create and annihilate a cavity mode excitation with wave-vector $k_z$; $\hbar\omega_{\text{exc}}(\mathbf R_j)$ is the excitation energy of molecule $j$, defined as: 
\begin{equation}
  \hbar\omega_{\text{exc}}(\mathbf R_j)=V_{\text{S}_1}^{\text{mol}}({\bf{R}}_j)-V_{\text{S}_0}^{\text{mol}}({\bf{R}}_j),
  \label{eq:Hnu}
\end{equation}
where $V_{\text{S}_0}^{\text{mol}}({\bf{R}}_j)$ and $V_{\text{S}_1}^{\text{mol}}({\bf{R}}_j)$ are the adiabatic potential energy surfaces (PESs) of molecule $j$ in the electronic ground (S$_0$) and excited (S$_1$) states, respectively. 
The last term in Equation~\ref{eq:dTCH} is the total potential energy of the system in the absolute ground state ({\it i.e.}, with all molecules and cavity modes de-excited), defined as the sum of the ground-state potential energies of all molecules in the cavity. The $V_{\text{S}_0}^{\text{mol}}({\bf{R}}_j)$ and $V_{\text{S}_1}^{\text{mol}}({\bf{R}}_j)$ adiabatic PESs are modelled at the hybrid quantum mechanics / molecular mechanics (QM/MM) level of theory\cite{Warshel1976b,Boggio-Pasqua2012}.

The third term in Equation~\ref{eq:dTCH} describes the light-matter interaction within the long-wavelength and rotating wave approximations:
\begin{equation}
    g_j(k_z) = -{\boldsymbol{\upmu}}({\bf{R}}_j) \cdot {\bf{u}}_{\text{cav}} \sqrt{\frac{\hslash\omega_{\text{cav}}(k_z)}{2\epsilon_0 V_{\text{cav}}}} 
    \label{eq:dipole_coupling}
\end{equation}
with ${\boldsymbol{\upmu}}({\bf{R}}_j)$ the transition dipole moment of molecule $j$ that depends on the molecular geometry (${\bf{R}}_j)$; ${\bf{u}}_\text{cav}$ the unit vector in the direction of the electric component of cavity vacuum field ({\it{i.e.}}, $|{\bf{E}}|=\sqrt{\hslash\omega_\text{cav}(k_z)/2\epsilon_0V_\text{cav}}$); $\epsilon_0$ the vacuum permittivity; and $V_{\text{cav}}$ the cavity mode volume.

\subsection{One-dimensional photonic crystal}

The distribution of the electric field in a one-dimensional distributed Bragg reflector (DBR), or photonic crystal (Figure~1 in the main text), with periodic dielectric constant, $\varepsilon(x)=\varepsilon(x+A)$ with period $A$, can be found by solving the wave equation~\cite{Sakoda2005}
\begin{equation}
\frac{c^2}{\varepsilon(x)}\frac{\partial^2E(x,t)}{\partial x^2}=-\frac{\partial^2E(x,t)}{\partial t^2},
   \label{eq:wave_equation1}
\end{equation}
where $c$ is the speed of light. Representing the electric field as a product of coordinate-dependent and time-dependent functions, $E(x,t)=E(x)e^{-i\omega t}$, leads to the following eigenvalue equation:
\begin{equation}
\frac{1}{\varepsilon(x)}\frac{\partial^2E(x)}{\partial x^2}=\frac{\omega^2}{c^2}E(x).
   \label{eq:wave_equation2}
\end{equation}

According to Bloch-Floquet theorem, a solution of Equation~\ref{eq:wave_equation2} can be chosen as 
\begin{equation}
   E_K(x) = u_K(x)e^{iKx},
   \label{eq:BF_theorem}
\end{equation}
where $u_k(x)=u_k(x+A)$ is a periodic envelope function and $K$ is the so-called Bloch wave number. For a Bloch surface wave, the Bloch wave number is complex and can be written as $K=i|K|$. Therefore, Equation~\ref{eq:BF_theorem} transforms into
\begin{equation}
   E_K(x) = u_K(x)e^{-|K|x}.
   \label{eq:BF_theorem_BSW}
\end{equation}

Thus, the electric field strength of the BSW is a decaying periodic function in the photonic crystal (black line for $x<0$ in Figure~1 in the main text). In the air, the electric field is not modulated by the periodicity of the dielectric structure and decays as (black line for $x>0$ in Figure~1 in the main text)
\begin{equation}
   E_K(x) = E_0e^{-|K|x}
   \label{eq:BF_theorem_BSW}
\end{equation}
with $E_0$ the electric field strength at the surface of the photonic crystal. 

While the electric field of the BSW is evanescent in the $x$-direction, there is no restriction on the field distribution along the surface of the DBR, \textit{i.e.} in the $y$- and $z$-directions. Therefore, light can freely propagate along the surface as a plane wave, and the electric field can be written as
\begin{equation}
 E(x,y,z)=E_0e^{-|K|x}e^{i(k_yy+k_zz)}.
   \label{eq:BSW_field}
\end{equation}

In the current work, we only consider a single layer of molecules and hence disregard 
the $x$-dependence in equation~\ref{eq:BSW_field}. Furthermore, we restrict ourselves to model one-dimensional transport along a chain of molecules in the $z$-direction. With these simplifications, the field distribution becomes 
\begin{equation}
    E(z)=E_0e^{ik_zz}.
   \label{eq:BSW_field_1D}
\end{equation}

Following Michetti and La Rocca,~\cite{Michetti2005} we impose periodic boundary conditions in the $z$-direction of the DBR and thus restrict the wave vectors, $k_z$, to discrete values: $k_{z,p}=2\pi p/L_z$ with $p \in \mathbb Z$ and $L_{z}$ the width of the DBR slab. With this approximation, the molecular Tavis-Cummings Hamiltonian in Equation~\ref{eq:dTCH} can be represented as an $(N+n_{\text{modes}})$ by $(N+n_{\text{modes}})$ matrix with four blocks\cite{Tichauer2021}:
\begin{equation}
  {\bf{H}}^{\text{TC}} = \left(\begin{array}{cc} 
  {\bf{H}}^{\text{mol}} & {\bf{H}}^{\text{int}}\\
  {\bf{H}}^{\text{int}\dagger} & {\bf{H}}^{\text{cav}}\\
  \end{array}\right).\label{eq:TavisCummings}
\end{equation}
We compute the elements of this matrix in the product basis of adiabatic molecular states times cavity mode excitations:
\begin{equation}
\begin{array}{ccl}
|\phi_j\rangle &=& \hat{\sigma}_j^+|\text{S}_0^1\text{S}_0^2..\text{S}_0^{N-1}\text{S}_0^N\rangle\otimes|00..0\rangle\\
\\
&=&\hat{\sigma}_j^+|\Pi_i^N\text{S}_0^i\rangle\otimes|\Pi_k^{n_\text{modes}}0_k\rangle \\
\\
&=&\hat{\sigma}_j^+|\phi_0\rangle
\end{array}\label{eq:basis1}
\end{equation}
for $1\le j\le N$, and 
\begin{equation}
\begin{array}{ccl}
|\phi_{j >N}\rangle &=& \hat{a}_{j-N}^\dagger|\text{S}_0^1\text{S}_0^2..\text{S}_0^{N-1}\text{S}_0^N\rangle\otimes|00..0\rangle\\
\\
&=&\hat{a}_{j-N}^\dagger|\Pi_i^N\text{S}_0^i\rangle\otimes|\Pi_k^{n_\text{modes}}0_k\rangle \\
\\
&=&\hat{a}_{j-N}^\dagger|\phi_0\rangle
\end{array}\label{eq:basis2}
\end{equation}
for $N < j\le N+n_\text{modes}$. In these expressions, $|00..0\rangle$ indicates that the Fock states for all $n_\text{modes}$ cavity modes are empty. The basis state $|\phi_0\rangle$ is the ground state of the molecule-cavity system with no excitations of neither the molecules nor cavity modes:
\begin{equation}
|\phi_0\rangle = |\text{S}_0^1\text{S}_0^2..\text{S}_0^{N-1}\text{S}_0^N\rangle\otimes|00..0\rangle
=|\Pi_i^N\text{S}_0^i\rangle\otimes|\Pi_k^{n_\text{modes}}0_k\rangle.\label{eq:phi0}
\end{equation}

The upper left block, ${\bf{H}}^{\text{mol}}$, is an $N\times N$ matrix containing the single-photon excitations of the molecules. Because we neglect direct excitonic interactions between molecules, this block is diagonal, with elements labelled by the molecule indices $j$:
\begin{equation}
    H_{j,j}^{\text{mol}}=\langle \phi_0|\hat{\sigma}_j\hat{H}^{\text{TC}}\hat{\sigma}_j^+|\phi_0\rangle\label{eq:molecular_diagonal}
\end{equation}
for $1 \le j \le N$. 
Each matrix element of ${\bf{H}}^\text{mol}$ thus represents the potential energy of a molecule, $j$, in the electronic excited state $|\text{S}_1^{j}({\bf{R}}_j)\rangle$, while all other molecules, $i\neq j$, are in the electronic ground state $|\text{S}_0^i({\bf{R}}_i)\rangle$:
\begin{equation}  H_{j,j}^{\text{mol}}=V_{\text{S}_1}^{\text{mol}}({\bf{R}}_j)+\sum^N_{i\neq j}V_{\text{S}_0}^{\text{mol}}({\bf{R}}_i).
  \label{eq:Hj}
\end{equation}

The lower right block,  ${\bf{H}}^{\text{cav}}$, is an $n_\text{modes}\times n_\text{modes}$ matrix (with $n_{\text{modes}}=n_{\text{max}}-n_{\text{min}}+1$) containing the single-photon excitations of the cavity (BSW) modes, and is also diagonal:
\begin{equation}
    H_{p,p}^{\text{cav}}= \langle \phi_0|\hat{a}_p\hat{H}^{\text{TC}}\hat{a}^\dagger_p|\phi_0\rangle\label{eq:photonic_diagonal}
\end{equation}
for $n_{\text{min}}\le p \le n_{\text{max}}$. Here, $\hat{a}_p^\dagger$ excites cavity mode $p$ with wave-vector $k_{z,p}= 2\pi p/L_z$. In these matrix elements, all molecules are in the electronic ground state $|\text{S}_0^i({\bf{R}}_i)\rangle$. The energy is therefore the sum of the cavity energy at $k_{z,p}$ and the molecular ground state energies:
\begin{equation}
    H_{p,p}^{\text{cav}}= \hslash\omega_{\text{cav}}(k_{z,p})+\sum^N_{j}V_{\text{S}_0}^{\text{mol}}({\bf{R}}_j),
    \label{eq:Hn}
\end{equation}
where $\omega_{\text{cav}}(k_{z,p})$ is the cavity dispersion (dashed-dotted curve in Figure 2\textbf{b}, main text). In this work, the dispersion of the BSW was fitted to reproduce the experimental dispersion from the study of Balasubrahmaniyam \textit{et. al.}~\cite{Balasubrahmaniyam2023}.

The two $N\times n_\text{modes}$ off-diagonal blocks ${\bf{H}}^{\text{int}}$ and ${\bf{H}}^{\text{int}\dagger}$ in the multi-mode Tavis-Cummings Hamiltonian (Equation~\ref{eq:TavisCummings}) model the light-matter interactions between the molecules and the cavity modes. Within the long-wavelength approximation these matrix elements can be approximated by the overlap between the transition dipole moment of molecule $j$ and the transverse electric field of cavity mode $p$ at the geometric center $z_j$ of that molecule:
\begin{equation}
\begin{array}{ccl}
  H_{j,p}^{\text{int}} &=& 
  -{\boldsymbol{\upmu}}({\bf{R}}_j)\cdot {\bf{u}}_{\text{cav}} \sqrt{\frac{\hslash\omega_{\text{cav}}(k_{z,p})}{2\epsilon_0 V_{\text{cav}}}} \langle\phi_0|\hat{\sigma}_j\hat{\sigma}^+_j\hat{a}_p e^{i 2\pi p z_j/L_z}\hat{a}_p^\dagger|\phi_0\rangle \\
  \\
  &=&
  -{\boldsymbol{\upmu}}({\bf{R}}_j) \cdot {\bf{u}}_{\text{cav}} \sqrt{\frac{\hslash\omega_{\text{cav}}(k_{z,p})}{2\epsilon_0 V_{\text{cav}}}} e^{i 2\pi p z_j/L_z}     
  \end{array}
  \label{eq:QMMMdipole_coupling}
\end{equation}
for $1 \le j \le N$ and $n_{\text{min}}\le p \le n_{\text{max}}$. 

\subsection{Ehrenfest dynamics}


In our simulations classical trajectories evolve under the influence of the expectation value of forces with respect to the polaritonic wave function\cite{Ehrenfest1927}, while the polaritonic wave function evolves along with the classical degrees of freedom. By expanding the total wave function as a linear combination of the \emph{time-independent} diabatic light-matter states (Equations~\ref{eq:basis1} and~\ref{eq:basis2}),
\begin{equation}
|\Psi(t)\rangle = \sum_j^{N+n_\text{modes}}|\phi_j\rangle d_j(t),
\label{eq:totalwf}
\end{equation}
the evolution of the \emph{time-dependent} diabatic expansion coefficients, $d_j(t)$, is obtained by numerically integrating the Schr\"{o}dinger equation over discrete time intervals, $\Delta t$, 
\begin{equation}
{\bf{d}}(t+\Delta t)={\bf{P}}^\text{dia}{\bf{d}}(t).
\label{eq:SE_dia}
\end{equation}
Here, ${\bf{d}}(t)$ is a vector containing the diabatic expansion coefficients $d_j(t)$ and ${\bf{P}}^\text{dia}$ is the propagator in the diabatic basis\cite{Sokolovskii2024b}:
\begin{equation}
    {\bf{P}}^\text{dia} =  \exp\left[-i\left({\bf{H}}^\text{TC}(t+\Delta t)+{\bf{H}}^\text{TC}(t)-i\hbar {\boldsymbol{\gamma}}\right)\Delta t/2\hbar\right],
    \label{eq:propdia}
\end{equation}
where ${\boldsymbol{\gamma}}$ is a vector containing decay rates $\gamma_j$ of each molecule and each cavity mode. Because in the current work, we simulate a short-time propagation of exciton-polaritons 
during two hundred femtoseconds, which is shorter than the lifetimes of both singlet excitons in Methylene Blue (MeB) molecule used in this study and Bloch surface waves\cite{Lerario2017,Forrest2020,Rashidi2024}, we neglect the decay rate vector ${\boldsymbol{\gamma}}$ in Equation~\ref{eq:propdia}, which results in
\begin{equation}
    {\bf{P}}^\text{dia} =  \exp\left[-i\left({\bf{H}}^\text{TC}(t+\Delta t)+{\bf{H}}^\text{TC}(t)\right)\Delta t/2\hbar\right].
    \label{eq:propdia_no_loss}
\end{equation}

\subsection{Resolving polariton transport in energy/momentum space}

To imitate an experimental approach\cite{Balasubrahmaniyam2023}, in which the transport of polaritonic states was resolved by probing their propagation at an energy and wave vector that match their locations along the lower BSW-polariton dispersion curve, we diagonalize the Tavis-Cummings Hamiltonian matrix (Equation~\ref{eq:TavisCummings}) at each time step of simulation to obtain \textit{adiabatic} polariton wave functions:
\begin{equation}
|\psi^m\rangle = \left(\sum_j^N\beta_j^m\hat{\sigma}_j^++\sum_p^{n_\text{modes}}\alpha_p^m\hat{a}^\dagger_p\right)|\phi_0\rangle=\sum_i^{N+n_\text{modes}}U_{im}|\phi_i\rangle
\label{eq:pol_states}
\end{equation}
with $\beta_j^m$ and $\alpha_p^m$ being the expansion coefficients, which reflect, respectively, the contribution of the molecular excitons and the cavity mode excitations to polariton state $\vert \psi^m\rangle$. Because the polariton states form a complete set, the total wave function in Equation~\ref{eq:totalwf} can be equivalently represented as a linear combination of these states:
\begin{equation}
|\Psi(t)\rangle=\sum_m^{N+n_{\text{modes}}}c_m(t)|\psi ^m\rangle,
\label{eq:totalwf_adia}
\end{equation}
Here, the time-dependent expansion coefficients, $c_m(t)$, associated with polariton states, are related to the diabatic expansion coefficients, $d_j(t)$ in Equation~\ref{eq:totalwf}, via the unitary transformation:
\begin{equation}
    {\bf{c}}(t) = {\bf{U}}^{-1}{\bf{d}}(t)={\bf{U}}^{\dagger}{\bf{d}}(t)
\end{equation}
with $\textbf{U}$ the unitary matrix that diagonalizes the Hamiltonian matrix and contains elements $U_{im}$ (Equation~\ref{eq:pol_states}).

To visualise polariton propagation, the probability density of the total wave function $|\Psi(t)|^2$ was constructed at each time step as a sum of excitonic $|\Psi_{\text{exc}}|^2$ and photonic $|\Psi_{\text{phot}}|^2$ contributions~\cite{Sokolovskii2023}. The amplitude of $|\Psi_{\text{exc}}(t)\rangle$ at position $z_j$ of molecule $j$ (with $z_j=(j-1)L_z/N$ for $1\le j \le N$) was obtained by projecting the excitonic basis state in which molecule $j$ at position $z_j$ is excited, onto the total wave function (Equation~\ref{eq:totalwf}):
\begin{equation}
    |\Psi_{\text{exc}}(z_j,t)\rangle=(\hat{\sigma}_j^+|\phi_0\rangle\langle\phi_0|\hat{\sigma}_j)|\Psi(t)\rangle = \sum_m^{N+n_\text{modes}}c_m(t)\beta_j^m\hat{\sigma}_j^+|\phi_0\rangle.
\label{eq:molecularpart}
\end{equation} 

The cavity mode excitations are described as plane waves that are delocalised in real space. We therefore obtained the amplitude of the cavity mode excitations in polaritonic eigenstate $|\psi^m\rangle$ at position $z_j$ by Fourier transforming the projection of the cavity mode Fock states, 
in which cavity mode $p$ is excited, onto $|\psi^m\rangle$: 
\begin{equation}
    |\psi^m_\text{phot}(z_j)\rangle=\mathcal{FT}^{-1}\left[\sum_p^{n_\text{modes}}(\hat{a}_p^\dagger|\phi_0\rangle\langle\phi_0|\hat{a}_p)|
    \psi^m\rangle\right] = \frac{1}{\sqrt N}\sum_p^{n_{\text{modes}}} \alpha_p^me^{ik_{z,p}z_j} \hat{a}_p^\dagger|\phi_0\rangle.
\label{eq:photonicpart1}
\end{equation}
The total contribution of the cavity mode excitations to the wavepacket at position $z_j$ at time $t$ was then obtained as the weighted sum over the Fourier transforms:
\begin{equation}
\begin{array}{ccl}
|\Psi_\text{phot}(z_j,t)\rangle &=& \sum_m^{N+n_\text{modes}}c_m(t)\times\mathcal{FT}^{-1}\left[\sum_p^{n_\text{modes}}(\hat{a}_p^\dagger|\phi_0\rangle\langle\phi_0|\hat{a}_p)|\psi^m\rangle\right]\\
\\
&=& \sum_m^{N+n_{\text{modes}}}c_m(t)\frac{1}{\sqrt N}\sum_p^{n_{\text{modes}}} \alpha_p^me^{ik_{z,p}z_j}\hat{a}_p^\dagger|\phi_0\rangle. 
\end{array}\label{eq:photonicpart2}
\end{equation}

The propagation of polaritonic states within a $k_z$-vector window, $w$, was obtained by constructing 
\textit{partial} excitonic and photonic wave functions, in which, rather than summing over all $N+n_{\text{modes}}$ polaritonic states, 
only lower polariton states with wave vectors lying in a certain window of width $\Delta k_z$ were included in the sum:
\begin{equation}
    |\Psi^{\text{part}}_{\text{exc},w}(z_j,t)\rangle=  \sum^{n_\text{modes}}_{m'\in w}c_{m'}(t)\beta_j^{m'}\hat{\sigma}_j^+|\phi_0\rangle.
\label{eq:Exc_wf_partial}
\end{equation}
and
\begin{equation}
|\Psi^{\text{part}}_{\text{phot},w}(z_j,t)\rangle=\frac{1}{\sqrt N}\sum^{n_\text{modes}}_{m'\in w}c_{m'}(t)\sum_p^{n_{\text{modes}}} \alpha_p^{m'}e^{ik_{z,p}z_j}\hat{a}_p^\dagger|\phi_0\rangle, 
\label{eq:Phot_wf_partial}
\end{equation}
for $\{m'|k_{z}^{m'}\in\left[k_z^c-\Delta k_z/2;k_z^c+\Delta k_z/2\right]\}$ with $k_z^c$ the wave vector corresponding to the centre of the window and $k_z^{m'}$ the expectation value of the in-plane momentum of polariton state $|\psi^{m'}\rangle$, evaluated as 
\begin{equation}
    \langle k_{z}^{m'} \rangle = \frac{\sum_p^{n_\text{modes}} |\alpha_p^{m'}|^2 k_{z,p} }{ \sum_p^{n_\text{modes}}|\alpha_p^{m'}|^2}
\label{eq:wave_vector_av}
\end{equation}
with $k_{z,p} = 2\pi p/L_z$. By restricting the sums over $m'$ to the first $n_{\text{modes}}=120$ eigenstates, the analysis only includes the states of the lower polaritonic branch. Finally, the probability amplitude of the partial wave function in window $w$ was computed as a sum of the probability amplitudes of the partial excitonic and photonic contributions:
\begin{equation}
    |\Psi_w^{\text{part}}(z,t)|^2 = |\Psi^{\text{part}}_{\text{exc},w}(z,t)|^2 + |\Psi^{\text{part}}_{\text{phot},w}(z,t)|^2.
\label{eq:WP_total_partial}
\end{equation}

As a measure of polariton transport, we calculated the mean squared displacement (MSD) of the partial wave function,
\begin{equation}
  \begin{array}{ccl}
    \text{MSD}_w(t)=\frac{\langle\Psi_w^{\text{part}}(z,t)|\left(\hat{z}(t)-\hat{z}(0)\right)^2|\Psi_w^{\text{part}}(z,t)\rangle}{\langle\Psi_w^{\text{part}}(z,t)|\Psi_w^{\text{part}}(z,t)\rangle},
  \end{array}
\label{eq:MSD_part}
\end{equation}
and of the partial photonic wave function,
\begin{equation}
  \begin{array}{ccl}
    \text{MSD}_{\text{phot},w}(t)=\frac{\langle\Psi_{\text{phot},w}^{\text{part}}(z,t)|\left(\hat{z}(t)-\hat{z}(0)\right)^2|\Psi_{\text{phot},w}^{\text{part}}(z,t)\rangle}{\langle\Psi_{\text{phot},w}^{\text{part}}(z,t)|\Psi_{\text{phot},w}^{\text{part}}(z,t)\rangle}.
  \end{array}
\label{eq:MSD_part_phot}
\end{equation}
To define the regime of polariton transport, the MSD was fitted with the following expression:
\begin{equation}
    \text{MSD}_w(t) = D_{\beta} t^\beta,
\label{eq:MSD_fit}
\end{equation}
where $D_{\beta}$ is a pre-factor and $\beta$ is the transport exponent that characterizes the propagation of the lower polaritonic states within $k_z$-vector window $w$. Subdiffusion is characterized by $\beta<1$; diffusion by $\beta=1$; super-diffusion by $1<\beta<2$; ballistic transport by $\beta=2$; and hyper-ballistic transport by $\beta>2$. 

\newpage

\section{Simulation details}

\subsection{Methylene Blue model}

\begin{figure*}[!htb]
\centering
\includegraphics[width=0.75\textwidth]{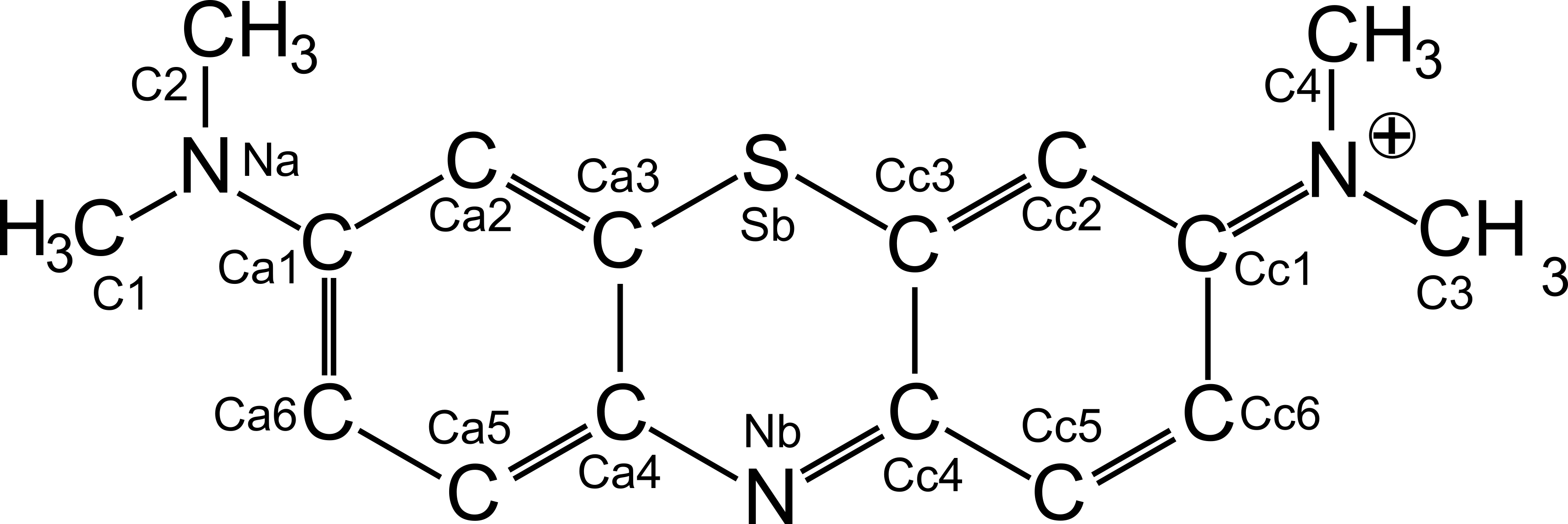}
  \caption{Structure and atom names of Methylene Blue. Not all hydrogens are shown, but their names take the name of the carbon atom plus an integer, {\it e.g.}, HA2 is attached to CA2 and H21, H22 and H23 are attached to carbon C2.}
\label{fig:MeB}
\end{figure*}

The Amber03 force field was used to model the interactions between MeB and the water solvent, which was described with the TIP3P water model\cite{Jorgensen1983}. Atom types and partial charges for the MeB atoms (Figure~\ref{fig:MeB}) are listed in Table~\ref{tab:MeB}. The partial charges of the atoms were derived following the procedure recommended for the Amber03~force field\cite{Duan2003,Bayly1993}. First, the geometry of MeB was minimized at the HF/6-31G** level of \textit{ab initio} theory, using the IEFPCM continuum solvent model with a relative dielectric constant of 4.0\cite{Tomasi2005}. After the geometry optimizations, the electrostatic potential at 10~concentric layers of 17~points per unit area around each atom was evaluated using the electron density calculated at the B3LYP/cc-pVTZ level of DFT theory\cite{Becke1993}, again using the IEFPCM continuum solvent model with a relative dielectric constant of 4.0\cite{Tomasi2005}. The atomic charges were obtained by performing a two-stage RESP fit to the electrostatic potential\cite{Bayly1993}, the first without symmetry constraints, and the second with symmetry constraints on chemically equivalent atoms.

\begin{table}[!htb]
\caption{Amber03 atomtypes and partial charges for Methylene Blue. Atom names are defined in Figure~\ref{fig:MeB}.}
\label{tab:MeB}
\begin{tabular}{l|l|r}
\hline
name  & type & charge (e)\\
\hline
  Ca1 & CA  &    0.054312 \\
   Na & NA  &    0.154719 \\
   C1 & CT  &   -0.284923 \\
  H11 & H1  &    0.128201 \\
  H12 & H1  &    0.128201 \\
  H13 & H1  &    0.128201 \\
   C2 & CT  &   -0.275632 \\
  H21 & H1  &    0.127095 \\
  H22 & H1  &    0.127095 \\
  H23 & H1  &    0.127095 \\
  Ca2 & CA  &   -0.205067 \\
  Ha2 & HA  &    0.181067 \\
  Ca3 & C   &   -0.101342 \\
  Ca4 & CA  &    0.624779 \\
  Ca5 & CA  &   -0.286158 \\
  Ha5 & HA  &    0.187369 \\
  Ca6 & CA  &   -0.141958 \\
  Ha6 & HA  &    0.158598 \\
   Nb & NB  &   -0.712421 \\
   Sb & S   &    0.049117 \\
  Cc1 & CA  &    0.054312 \\
   Nc & NA  &    0.154719 \\
   C3 & CT  &   -0.284923 \\
  H31 & H1  &    0.128201 \\
  H32 & H1  &    0.128201 \\
  H33 & H1  &    0.128201 \\
   C4 & CT  &   -0.275632 \\
  H41 & H1  &    0.127095 \\
  H42 & H1  &    0.127095 \\
  H43 & H1  &    0.127095 \\
  Cc2 & CA  &   -0.205067 \\
  Hc2 & HA  &    0.181067 \\
  Cc3 & C   &   -0.101342 \\
  Cc4 & CA  &    0.624779 \\
  Cc5 & CA  &   -0.286158 \\
  Hc5 & HA  &    0.187369 \\
  Cc6 & CA  &   -0.141958 \\
  Hc6 & HA  &    0.158598 \\
\hline
\end{tabular}
\end{table}

A single MeB molecule was geometry-optimized at the B97/3-21G level of DFT theory and placed at the center of a rectangular box that was filled with 2031 TIP3P water molecules\cite{Jorgensen1983}. A 1.0~nm cut-off was used for the Van der Waals' interactions, which were modeled with Lennard-Jones potentials, while the Coulomb interactions were computed with the smooth particle mesh Ewald method\cite{Essmann1995}, using a 1.0~nm real space cut-off and a grid spacing of 0.12~nm and a relative tolerance at the real space cut-off of 10$^{-5}$. The simulation box, containing 6,131 atoms, was equilibrated for 1~ns at the force field level of theory. During this equilibration, the coordinates of the MeB atoms were kept fixed. The temperature was maintained at 300~K with the v-rescale thermostat\cite{Bussi2007}, while the pressure was kept constant at 1~atmosphere using the Berendsen isotropic pressure coupling algorithm\cite{Berendsen1984}, with a time constant of 1~ps. The SETTLE algorithm was applied to constrain the internal degrees of freedom of water molecules\cite{Miyamoto1992}, enabling a time step of 2~fs in the classical MD simulations. 

After equilibration at the force field (MM) level, the system was further equilibrated at the QM/MM level for 10~ps. The time step was reduced to 1~fs. The Methylene Blue molecule was modelled at the DFT level, using the B97 functional\cite{Becke97}, in combination with the 3-21G basis set\cite{Dunning1970}. The water solvent was modelled with the TIP3P force field\cite{Jorgensen1983}. The QM system experienced the Coulomb field of all MM atoms within a 1.0~nm cut-off sphere and Lennard-Jones interactions between MM and QM atoms were added. The singlet electronic excited state (S$_1$) was modeled with time-dependent DFT (TD-DFT)\cite{Runge1984}, using the B97 functional in combination with the 3-21G basis set for the QM region ({\it {i.e.,}} TD-B97/3-21G)\cite{Dunning1970}. At this level of QM/MM theory, the excitation energy is 2.5~eV. The overestimation of the vertical excitation energy is due to the limited accuracy of the employed level of theory, but it can be easily compensated by adding an off-set to the cavity resonance energy. The QM/MM simulations were performed with GROMACS version 4.5.3\cite{Hess2008}, interfaced to Gaussian16\cite{g16}.

\subsection{MeB-BSW system}
\label{subsec:cavitymd}

\begin{figure*}[!tb]
\centering
\includegraphics[width=1\textwidth]{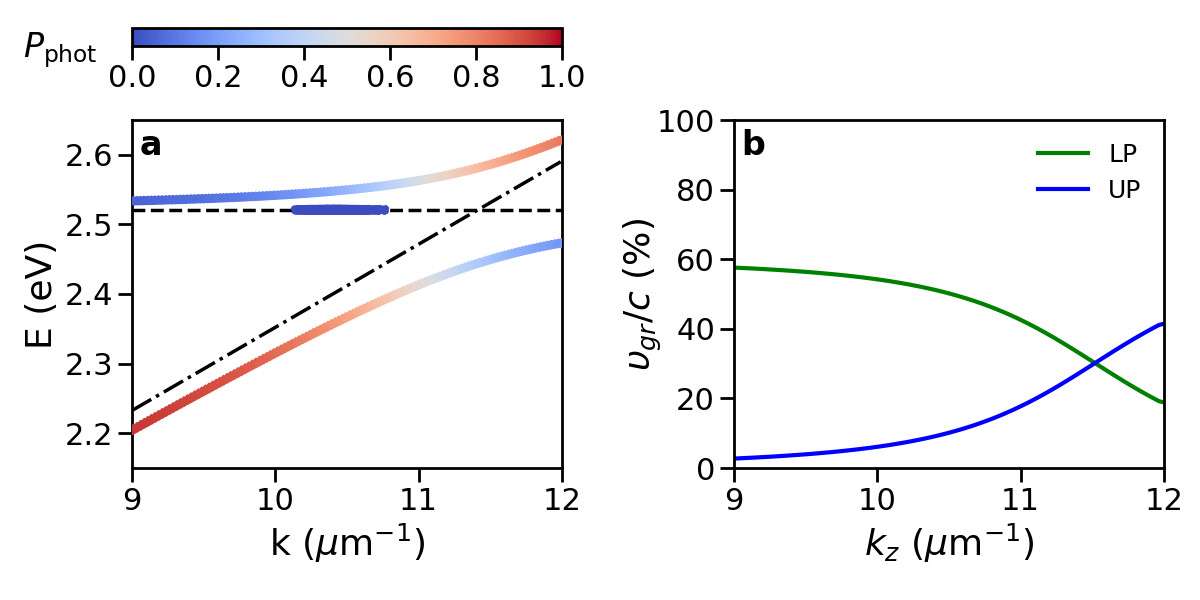}
  \caption{Panel \textbf{a}: Dispersion of BSW-polaritons. The total contribution of all cavity modes to each polaritonic state ($P_{\text{phot}}$) is indicated by colors. Dispersion of the BSW is shown as the dashed-dotted line, and the first excited state of the initial configuration of Methylene Blue is shown as the dashed line. Panel \textbf{b}: Group velocity, $\upsilon_{\text{gr}}=\partial\omega_{\text{pol}}/\partial k_z$, of lower (LP, green line) and upper (UP, blue line) polaritons as a fraction of the speed of light, $c$.}
\label{fig:dispersion}
\end{figure*}

From a QM/MM trajectory of MeB in the S$_0$ state, a single snapshot with an $S_0\rightarrow S_1$ excitation energy of $E_{\text{exc}}=2.52$~eV (dashed line in Figure~\ref{fig:dispersion}\textbf{a}) was selected as the initial configuration for each of $N=1024$ molecules in the system. These molecules, including their solvent environment, were placed along the $z$-axis at equal intermolecular separations on the surface of a one-dimensional photonic crystal (Figure~1 in the main text) of width $L_z=250~\mu$m. The experimental dispersion of the Bloch surface wave~\cite{Balasubrahmaniyam2023} was fitted with a linear function, $E_{\text{BSW}}(k_z)=a\cdot k_z+b$ with $a=0.119$~eV$\mu$m$^{-1}$ and $b=0.771$~eV, and is shown as the dashed-dotted line in Fugure~\ref{fig:dispersion}\textbf{a}. Additionally, an offset $\Delta E=0.37$~eV was added to the energy of the BSW to compensate for the energy difference between the excitation energy $E_{\text{exc}}^{\text{TDBC}}=2.13$~eV of TDBC J-aggregates used in the experiment~\cite{Balasubrahmaniyam2023} and the absorption maximum $E_{\text{exc}}^{\text{MeB}}=2.50$~eV of MeB at the TD-B97/3-21G level of TDDFT theory. The dispersion of the BSW was modelled with 120~discrete modes ({\it i.e.}, $k_{z,p}=2\pi p/L_z$ with $358\leq p\leq 477$). To maximize the collective light-matter coupling strength, the transition dipole moments of all molecules were aligned to the vacuum field at the start of the simulation. With a vacuum field strength of ${\bf{E}}_y =$ 0.000014022 a.u. (0.071~MV~cm$^{-1}$) the Rabi splitting, defined as the energy gap between the upper (UP) and lower (LP) polaritons at the wave vector where the molecular excitation energy matches the cavity mode energy, was 131~meV, close to the Rabi splitting of 142~meV in the experimental study\cite{Balasubrahmaniyam2023}. The resulting BSW-polariton dispersion is depicted in Figure~\ref{fig:dispersion}\textbf{a} with colors indicating the contribution of all BSW modes to each polariton state. In addition, the group velocities of the LP and UP branches are shown in Figure~\ref{fig:dispersion}\textbf{b}.

All simulations were inititated with an off-resonant excitation into the first excited state, S$_1$, of the MeB molecule at the centre of the molecular chain, \textit{i.e.}, at $z=125~\mu$m. To analyse how polariton transport changes along the LP branch, partial wave functions, $|\Psi^{\text{part}}(z,t)\rangle$ (Equations~\ref{eq:Exc_wf_partial},~\ref{eq:Phot_wf_partial} and~\ref{eq:WP_total_partial}), were extracted from wave vector windows of width $\Delta k_z=0.5~\mu$m$^{-1}$ centred at different wave vectors, $k_z^c$, according to Table~\ref{tab:windows}. 

\begin{table}[!htb]
\caption{Ranges of wave vectors ($k_z$, first column) at which partial wave functions, $|\Psi^{\text{part}}(z,t)\rangle$, were extracted. The second and third columns indicate, respectively, the wave vector of the polariton state corresponding to the centre of each window ($k_z^c$) and the total contribution of all BSW modes ($|\alpha_m^c|^2=\sum_p^{n_{\text{modes}}}|\alpha_p^m|^2$) to this state.}
\label{tab:windows}
\begin{tabular}{c|c|c}
\hline
  $k_z$-range ($\mu$m$^{-1}$) & $k_z^c$ ($\mu$m$^{-1}$) & $|\alpha_m^c|^2$ \\
  \hline
  $\left[9.00; 9.50\right]$ & 9.25 & 0.93 \\
  $\left[9.25; 9.75\right]$ & 9.50 & 0.91 \\
  $\left[9.50; 10.00\right]$ & 9.75 & 0.89 \\
  $\left[9.75; 10.25\right]$ & 10.00 & 0.85 \\
  $\left[10.00; 10.50\right]$ & 10.25 & 0.80 \\
  $\left[10.25; 10.75\right]$ & 10.50 & 0.73 \\
  $\left[10.50; 11.00\right]$ & 10.75 & 0.64 \\
\hline
\end{tabular}
\end{table}

\clearpage

\section{Analysis of the wave packets}

\subsection{Simulation of $N=1024$ molecules}
\label{sec:1024mol}

\subsubsection{Partial wave packet}
\label{sec:total_WP}

\begin{figure}[!htb]
\centering
\includegraphics[width=1\textwidth]{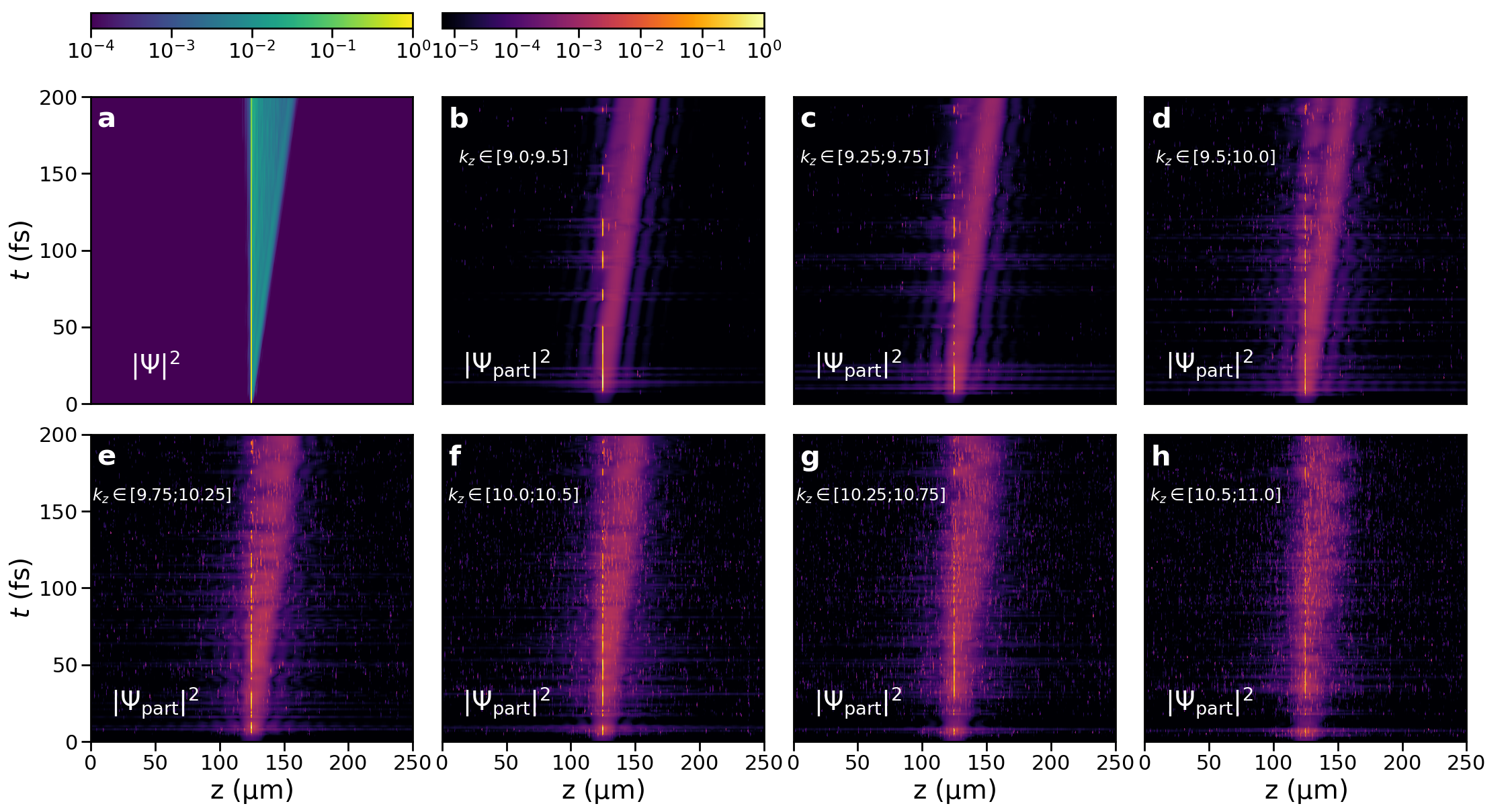}
  \caption{Time-space maps of the probability amplitude of the total polariton wave function $|\Psi|^2$ (panel \textbf{a}) and of the partial wave function $|\Psi_w^{\text{part}}|^2$ (panels~\textbf{b}-\textbf{h}) associated with different wave vector windows (Table~\ref{tab:windows}).}
  \label{fig:BSW_transport}
\end{figure}

Figure~\ref{fig:BSW_transport}\textbf{a} shows a space-time map of the probability amplitude of the total wave function ($|\Psi(z,t)|^2$, Equation~\ref{eq:totalwf}). After initial excitation into the molecule located at $z=125~\mu$m on the DBR surface, a wave packet of polaritons forms and spreads in the positive direction along the $z$-axis on a micrometer scale. To extract how different lower polaritonic states with well-defined wave vectors, contribute to the propagation of the total wavefunction, we constructed  partial wave functions, $|\Psi_w^{\text{part}}(z,t)\rangle$ (Equations~\ref{eq:Exc_wf_partial},~\ref{eq:Phot_wf_partial} and~\ref{eq:WP_total_partial}), in windows $w$, representing a fixed wave vector interval (Table~\ref{tab:windows}). The time-evolution of the probability densities of the polaritonic states within these windows ($|\Psi_w^{\text{part}}(z,t)|^2$, Equation~\ref{eq:WP_total_partial}), is shown in Figure~\ref{fig:BSW_transport}\textbf{b}--\textbf{h}. In all windows, the partial wave packet formed by the states within the window, propagates, indicating that all LP states with a well-defined wave vector, contribute to the overall transport.

\begin{figure}[!htb]
\centering
\includegraphics[width=0.7\textwidth]{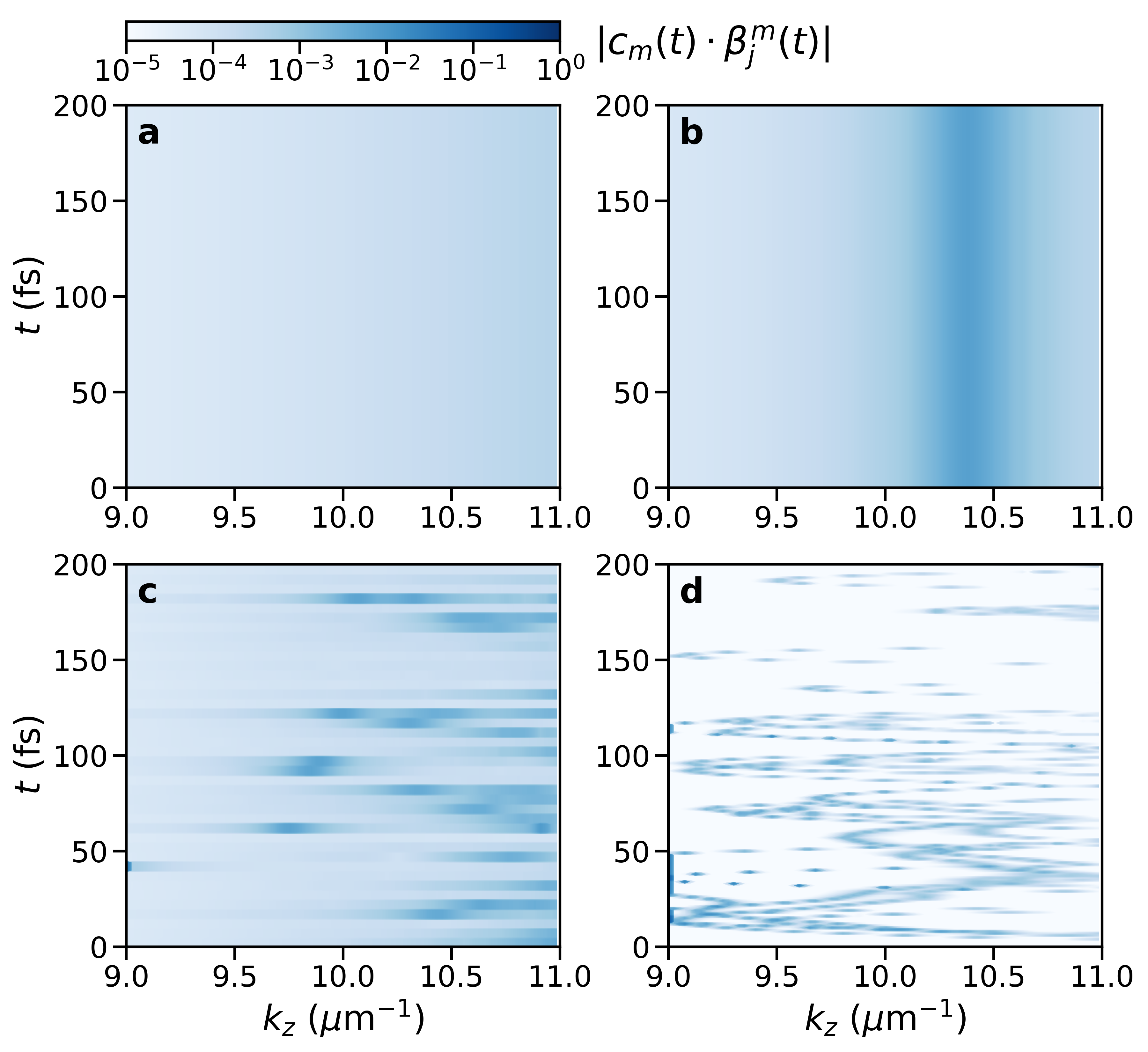}
  \caption{Contributions of the initially excited molecule $j$ at time $t$ to lower polaritonic states $|\psi_m\rangle$ with wave vectors between $k_z=9$~and~$11~\mu$m$^{-1}$ (\textit{i.e.}, $|c_m(t)\cdot \beta_j^m(t)|$). Panels \textbf{a} and \textbf{b}: Simulations of two-level systems without ($\sigma=0$~meV, \textbf{a}) and with ($\sigma=74$~meV, \textbf{b}) static disorder among the molecular excitation energies. Panel~\textbf{c}: Simulations with quasi-dynamical disorder, modelled by resampling the excitation energies of the two-level systems every five femtoseconds. Panel~\textbf{d}: QM/MM molecular dynamics simulations. The plot in panel~\textbf{a} corresponds to a single run, the plots in panels~\textbf{b} and \textbf{c} are averages of 25 runs, and the plot in panel~\textbf{d} is averaged over five runs.}
  \label{fig:beta_contr}
\end{figure}

In addition to the propagating part of the partial wave function, a fraction of the population remains in the initially excited molecule, which is manifested by the vertical line at $z=125~\mu$m in Figure~\ref{fig:BSW_transport}\textbf{b}--\textbf{h}. This line, however, appears and reappears at different time points, depending on the window from which the partial wave function was extracted. To understand the origin of this observation, we extracted the contribution of the initially excited molecule $j$ at time $t$ to all lower polariton states within the wave vector range of $9-11~\mu$m$^{-1}$, and plotted these contributions as a function of $k_z$ in Figure~\ref{fig:beta_contr}. These contributions were thus obtained as the total amplitude of the excitation of molecule $j$ to each of the eigenstates $|\Psi_m\rangle$ on the lower polariton branch: $|c_m(t)\beta_j^m(t)|$. Here, the 80 lowest-energy eigenstates were considered to form the LP branch, as these states are well separated from the dark state manifold and therefore have well-defined $k_z$-vectors. The expectation value of the $k_z$-vector for each of these states, $|\Psi_m\rangle$, was determined as $\langle k_z^m\rangle = \sum_p^{n_\text{modes}}|\alpha_p^m|^2 k_{z,p}/ \sum_p^{n_\text{modes}}|\alpha_p^m|^2$.


First, we analyzed the contribution of the excitation of the initially excited molecule to the LP states, when the molecules were modeled as static two-level systems without disorder. As shown in Figure~\ref{fig:beta_contr}\textbf{a}, in such simulations, the contribution remains constant over time, but increases with $k_z$, as at higher $k_z$ values, the LP branch approaches the molecular excitation energy, and polariton states become more exciton-like\cite{Agranovich2003,Litinskaya2004}. When we included static disorder of the excitation energies, the variation of $|c_m(t)\beta_j^m(t)|$ with $k_z$ is also constant in time, but no longer monotonously increases with $k_z$, and instead has a maximum around $k_z = 10.4~\mu$m$^{-1}$ (Figure~\ref{fig:beta_contr}\textbf{b} for $\sigma=74$~meV). In contrast, in simulations of two-level systems with quasi-dynamic disorder, the contribution of the initial molecular excitation to the LP states continuously re-distributes among the LP states  (Figure~\ref{fig:beta_contr}\textbf{c}). In these simulations, the excitation energies of the two-level systems were resampled every five femtoseconds and the dynamics were averaged over 25 individual runs\cite{Nitzan2023}. The time-dependent distribution of $|c_m(t)\beta_j^m(t)|$ in these simulations is qualitatively similar to the time-dependent distribution extracted from the atomistic QM/MM MD simulations (Figure~\ref{fig:beta_contr}\textbf{d}). We, therefore, attribute the vertical lines that appear and reappear at $z=125~\mu$m in the plots of the partial wave functions in different windows (Figure~\ref{fig:BSW_transport}\textbf{b}--\textbf{h}), to dynamical disorder, which leads to a continuous redistribution of the excitation energies, and hence to a time-dependent fluctuation of the contribution of the initially excited molecule to the LP states.


\begin{figure}[!htb]
\centering
\includegraphics[width=1\textwidth]{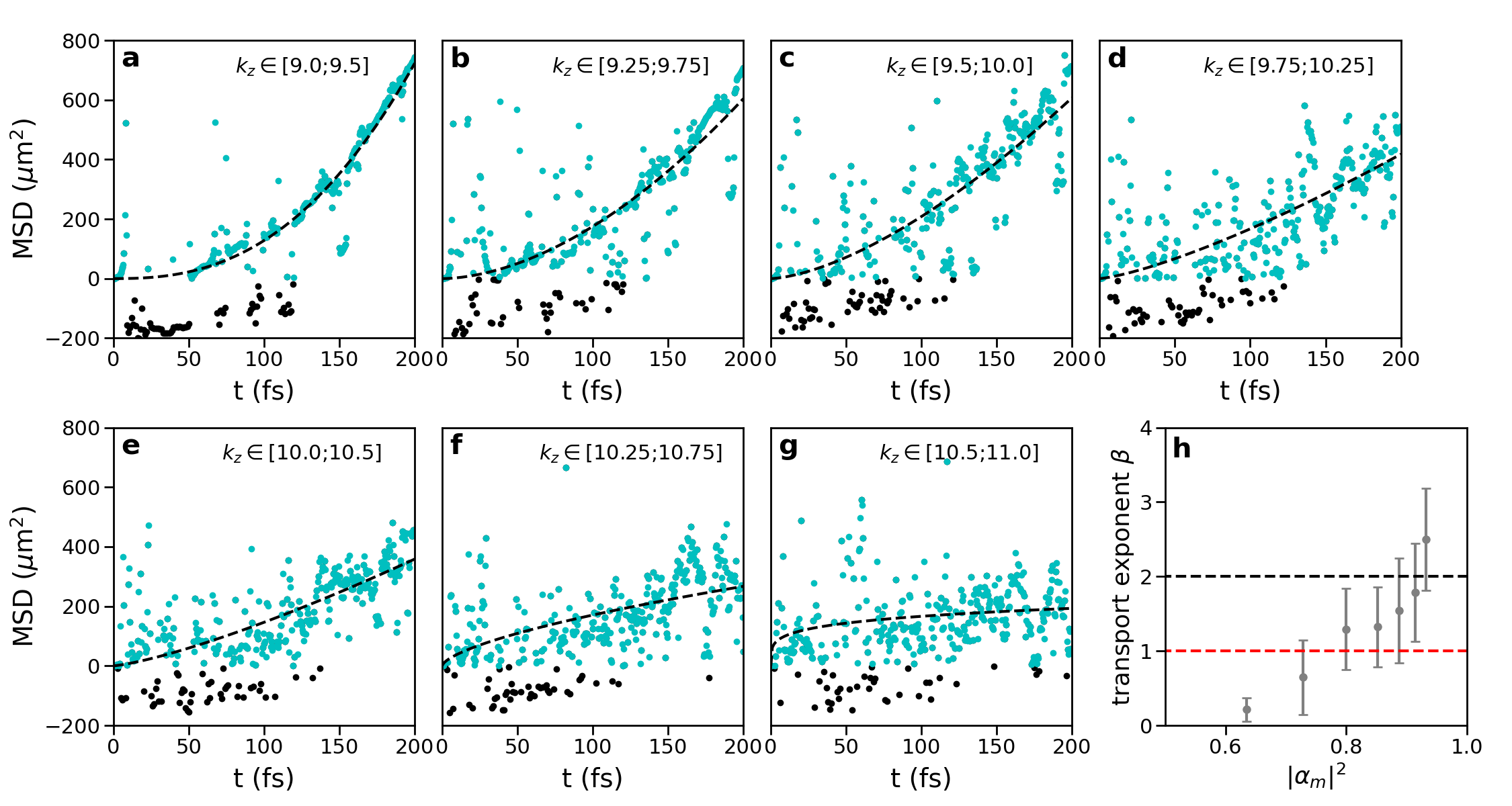}
  \caption{Panels~\textbf{a}-\textbf{g}: Mean squared displacement, $\text{MSD}_w(t)-\text{MSD}_w(0)$, of partial wave functions $|\Psi_{\text{part}}|^2$ as a function of time, extracted from different wave vector windows. Dashed lines are fits of the cyan dots to $D_{\beta}\cdot t^\beta$. Panel~\textbf{h}: The transport exponent $\beta$ as a function of the BSW modes contribution $|\alpha_{\text{m}}|^2$ to polaritonic states. The error bars are standard deviations of five runs.}
  \label{fig:MSD_tot}
\end{figure}

To quantify how the transport regime of polaritons changes along the LP branch, we computed the mean squared displacements (MSD$_w$, Equation~\ref{eq:MSD_part}) of the partial wave functions associated with the windows (Figure~\ref{fig:MSD_tot}\textbf{a}-\textbf{g}), and fitted these MSD$_w$'s to Equation~\ref{eq:MSD_fit} in order to extract the transport exponents, $\beta$. The time-dependent fluctuations of the contributions of the molecular excitations to the LP states, discussed above, make the MSD$_w$ plots noisy, which is manifested by MSD$_w$ values below zero (black dots). These data points were therefore excluded from the fits, which were thus performed with only the cyan dots. The fits of Equation~\ref{eq:MSD_fit} to these cyan points are shown as black dashed lines.


In Figure~\ref{fig:MSD_tot}\textbf{h}, the transport exponent, $\beta$, is plotted as a function of the total contribution of all BSW modes, $|\alpha^w_m|^2$, to the state ($|\psi_m^w\rangle$) whose wave vector is at the centre of the window $w$ (Table~\ref{tab:windows}). These plots suggest that the transport exponent increases with the photonic content as we move down in energy along the LP branch.
The decrease of the transport exponent when the photonic content of a LP state decreases, indicates a transition of the transport regime from ballistic motion at large photonic contents to diffusion at lower photonic contents, in line with experiments\cite{Xu2022,Balasubrahmaniyam2023}. Quantitatively, the value of $\beta$ decreases from $\sim2.4$ at $|\alpha^w_m|^2=0.93$ to $\sim0.2$ at $|\alpha^w_m|^2=0.64$. The latter value is characteristic of subdiffusion, which was not observed in the experiment\cite{Balasubrahmaniyam2023}. This may be due to the fact that the polariton states with wave vectors belonging to the windows $k_z\in \left[10.25; 10.75\right]~\mu$m$^{-1}$ and $k_z\in \left[10.50; 11.00\right]~\mu$m$^{-1}$ are rather close in energy to the molecular excitation energies , 
and may hence be "contaminated" by molecular excitations that are at the lower energy part of the molecular absorption spectrum\cite{Agranovich2003,Litinskaya2004}.
Although by increasing the number of simulations, we may average out the impact of these molecular excitations on the transport coefficient, our current computational resources did not allow us to go beyond five trajectories. Nevertheless, despite the very large statistical errors that are unavoidable for such a small number of trajectories, we observe a clear trend that is furthermore in good qualitative agreement with the experiments~\cite{Balasubrahmaniyam2023,Xu2022}.

\subsubsection{Partial photonic wave packet}

\begin{figure}[!htb]
\centering
\includegraphics[width=1\textwidth]{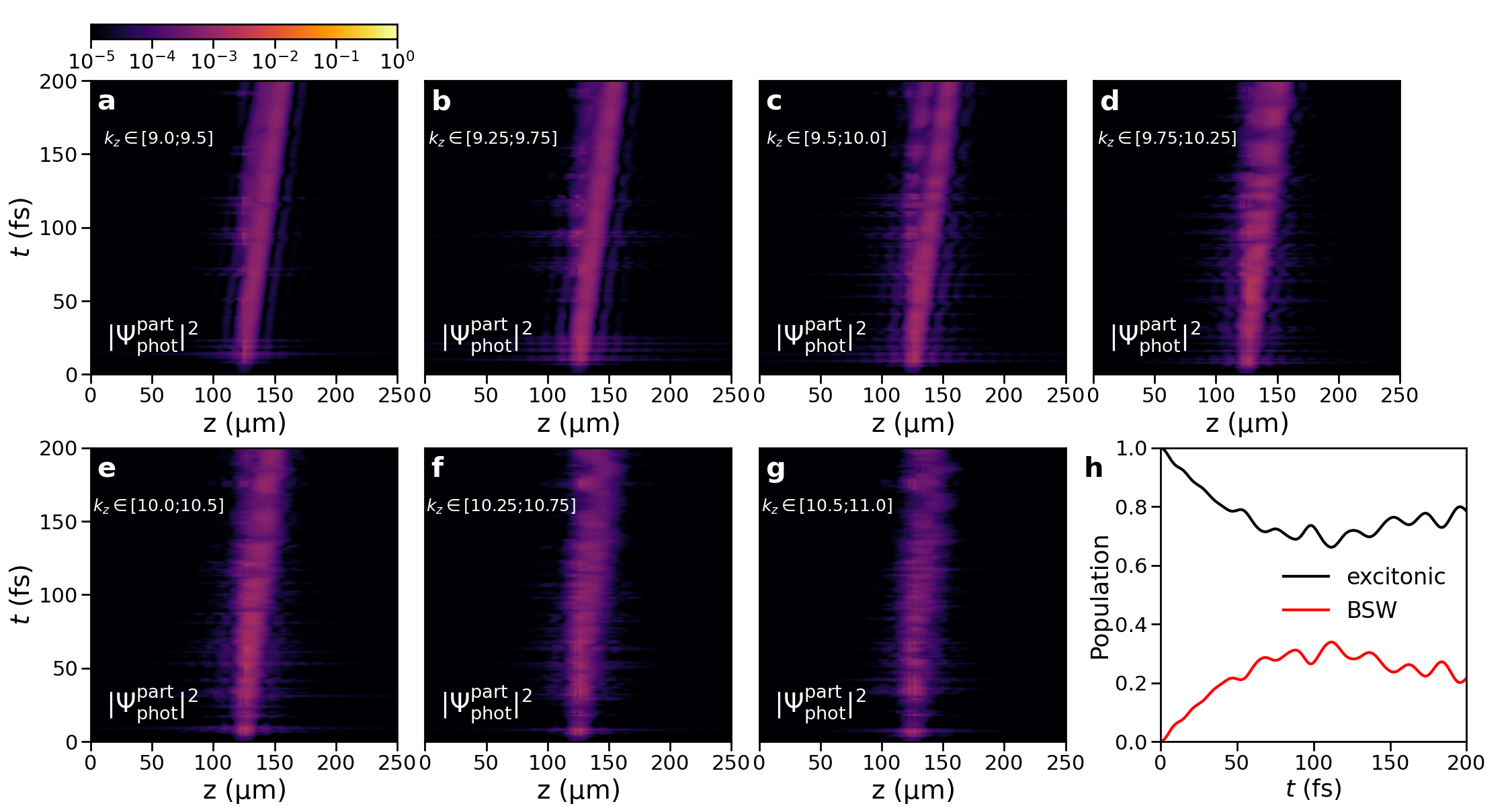}
  \caption{Panels~\textbf{a}-\textbf{g}: Time-space maps of the probability amplitude of the partial photonic wave function, $|\Psi^{\text{part}}_{\text{phot},w}(z,t)|^2$, associated with different wave vector windows (Table~\ref{tab:windows}).
  Panel~\textbf{h}: Contributions of molecular excitons (black) and BSW excitations (red) to $|\Psi(z,t)|^2$.}
  \label{fig:BSW_transport_phot}
\end{figure}

Because the photonic part of the wave function is less sensitive to the fluctuations of the molecular excitation energies than the total, or excitonic wave functions, we focus our analysis on the partial \textit{photonic} wave functions ($|\Psi^{\text{part}}_{\text{phot},w}(z,t)\rangle$, Equation~\ref{eq:Phot_wf_partial}) instead, and  avoid the large statistical fluctuations in our estimates of the transport coefficient.
In Figure~\ref{fig:BSW_transport_phot}\textbf{a}-\textbf{g} we show the probability amplitude of these partial photonic wave functions, $|\Psi^{\text{part}}_{\text{phot},w}(z,t)|^2$, extracted from the same wave vector windows as in Section~\ref{sec:total_WP}. Comparing these plots to the plots in Figure~\ref{fig:BSW_transport}, we indeed find that the effect of the fluctuations in the molecular excitation energies is much smaller when we restrict our analysis to the photonic part of the wave functions than when we analyze partial wave functions that contain both the molecular and photonic parts. 

\begin{figure}[!htb]
\centering
\includegraphics[width=1\textwidth]{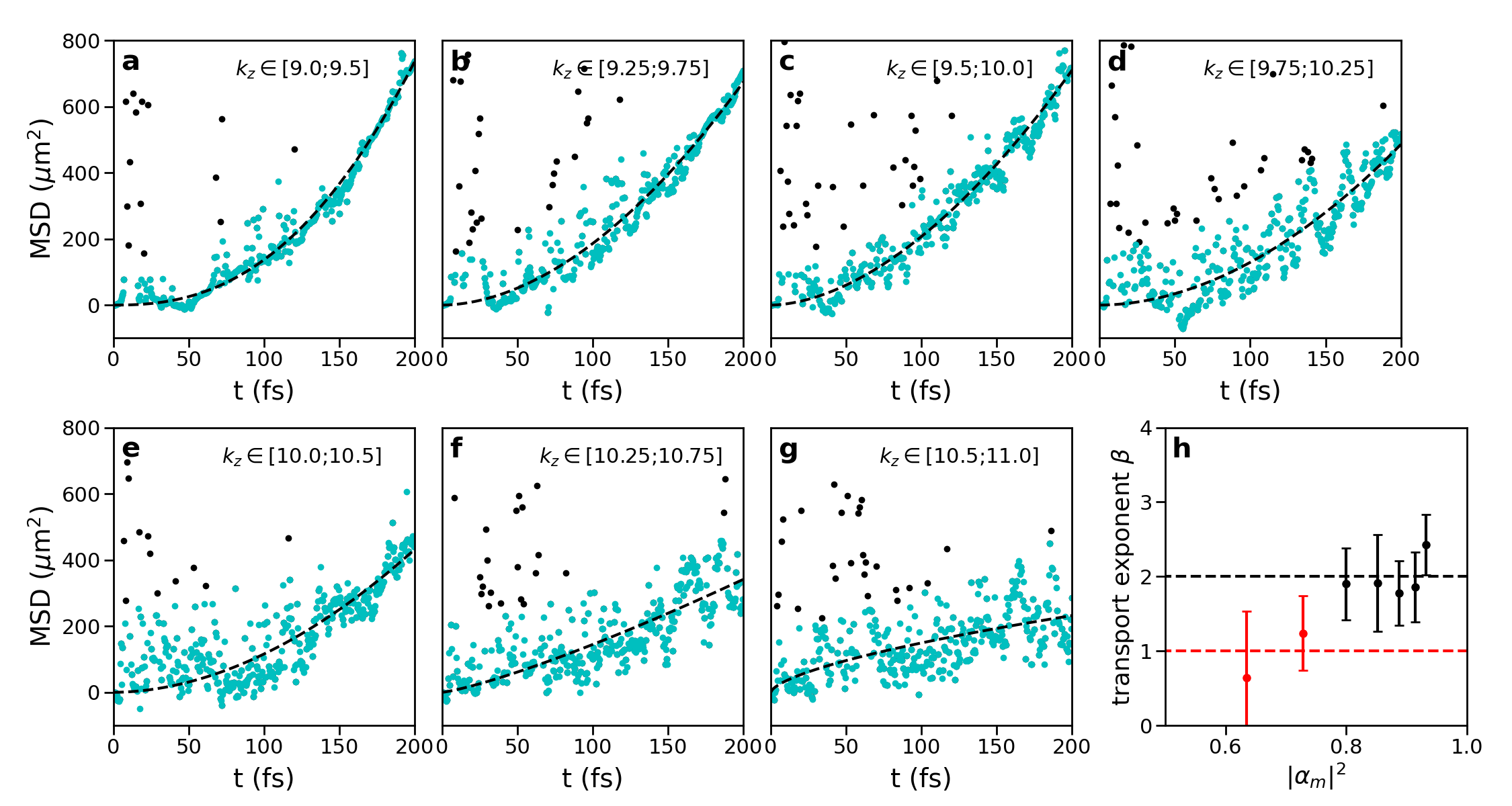}
  \caption{Panels~\textbf{a}-\textbf{g}: Mean squared displacement, $\text{MSD}_w(t)-\text{MSD}_w(0)$, of partial photonic wave functions $|\Psi^{\text{part}}_{\text{phot},w}|^2$ as a function of time, extracted from different wave vector windows. Dashed lines are fits of $D_{\beta}\cdot t^\beta$ to the cyan dots. Panel~\textbf{h}: The transport exponent $\beta$ as a function of the BSW modes contribution $|\alpha_{\text{m}}|^2$ to polaritonic states. The error bars are standard deviations of five runs.}
  \label{fig:MSD_part_phot}
\end{figure}

In Figure~\ref{fig:MSD_part_phot}\textbf{a}-\textbf{g}, we plot the MSD$_w$ of the partial photonic wave functions, as well as fits to $D_{\beta}\cdot t^{\beta}$ (Equation~\ref{eq:MSD_fit}). As before, when fitting the MSD$_w$'s for the partial wave functions containing both excitonic and photonic parts, we had to omit some points, shown in black, from these fits. However, in contrast to the previous situation, we here attribute the outliers to artefacts introduced by the Fourier transformation of the photonic wave functions into real space. These artefacts are visible as 
horizontal stripes in the time-space maps of $|\Psi_w^{\text{part}}(z,t)|^2$ in Figure~\ref{fig:BSW_transport_phot}\textbf{a}-\textbf{g}, and lead to an overestimation of the MSD$_w$ values. We thus only used the data points shown in cyan to fit the MSD expression (Equation~\ref{eq:MSD_fit}) and obtain the transport exponents.


The transport exponents associated with the partial photonic wave functions in the different $k_z$-vector windows (Figure~\ref{fig:MSD_part_phot}\textbf{h}), show a similar trend as the ones associated with $|\Psi_w^{\text{part}}(z,t)|^2$ and suggest a crossover between ballistic transport and diffusion. 
Compared to the analysis of the partial wave function containing both the excitonic and photonic contributions (Figure~\ref{fig:MSD_tot}{\bf{h}}), the transport coefficients obtained for the photonic part of the wave packet, have smaller error bars and rise less steeply with $k_z$. 
Furthermore, in line with the experiment\cite{Balasubrahmaniyam2023}, for all ranges of well-defined wave vectors, the transport exponent remains, within the error, equal to one or two, with a crossover between the two transport regimes around a photonic content of $|\alpha_m|^2 \approx 0.7$. 

\clearpage

\subsection{Simulation of two-level systems with static disorder}

\begin{figure}[!htb]
\centering
\includegraphics[width=1\textwidth]{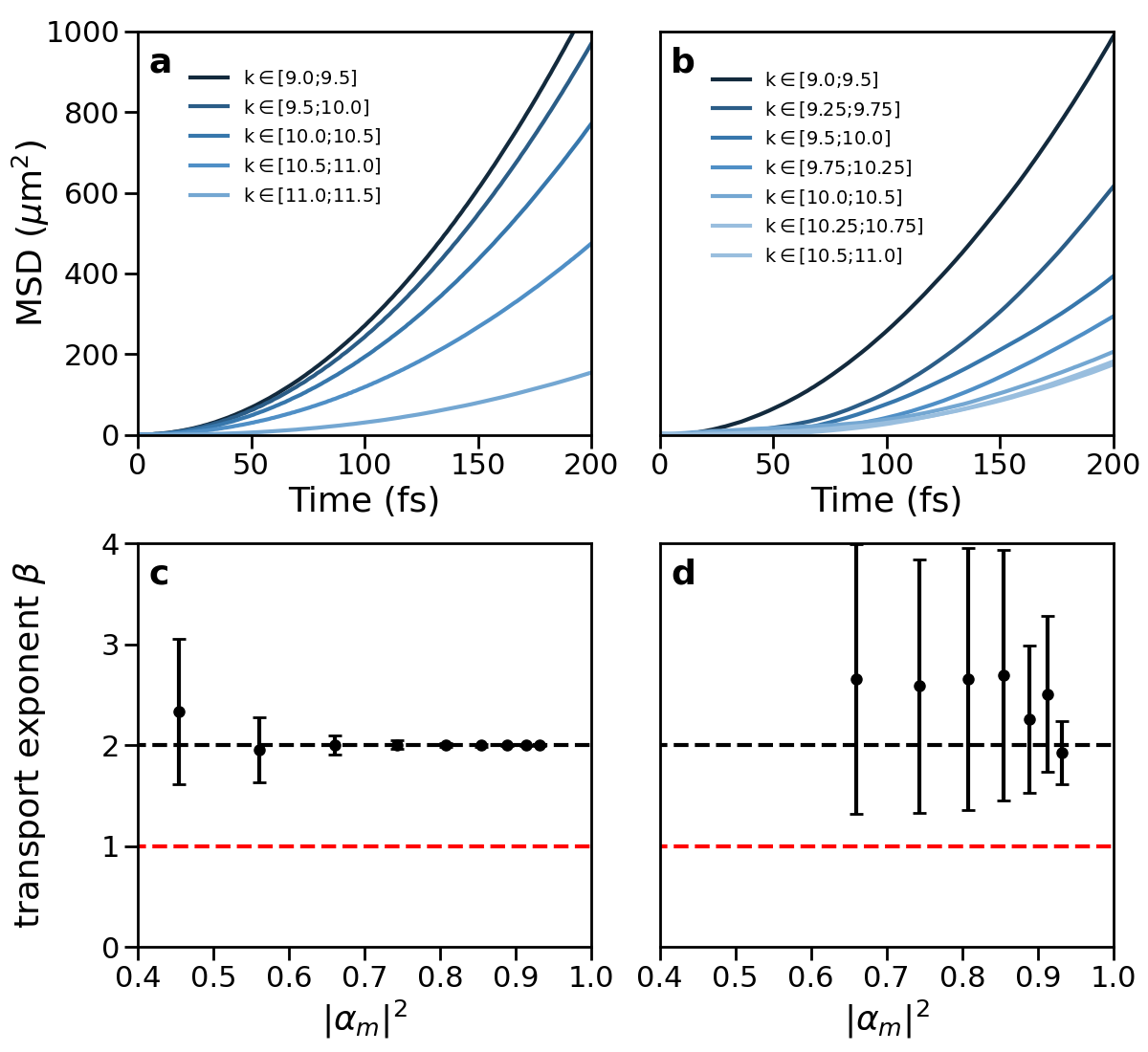}
  \caption{Panels~\textbf{a}~and~\textbf{b}: Mean squared displacement (MSD$_w$) of the partial photonic wave function $|\Psi^{\text{part}}_{\text{phot}}|^2$ extracted from different wave vector windows in 
  static simulations of two-level systems with excitation energy disorder of $\sigma=22$~meV (\textbf{a}) and $\sigma=63$~meV (\textbf{b}). 
  Panels~\textbf{c}~and~\textbf{d}: Values of the transport exponent, $\beta$, as a function of the BSW modes contribution $|\alpha_{\text{m}}|^2$ to polaritonic states. The errors are standard deviations of five hundred individual runs.}
  \label{fig:transport_static}
\end{figure}

To understand whether the transition from ballistic to diffusive transport along the lower polariton branch is caused by population exchanges between polariton states due to nonadiabatic coupling or whether it can be solely explained by excitation energy disorder, we performed simulations of $N=1000$ two-level systems with 
static disorder. In these simulations, the excitation energies of the two-level systems were drawn from a Gaussian distribution (Equation~9 in the main text). In Figure~\ref{fig:transport_static}\textbf{a}-\textbf{b} we show the mean squared displacement of the partial photonic wave function $|\Psi^{\text{part}}_{\text{phot},w}|^2$ extracted from different wave vector windows in simulations with disorder strengths of $\sigma=22$~meV and $\sigma=63$~meV, respectively. These values correspond to the line-width of the absorption spectra of TDBC J-aggregates used in the experiment\cite{Balasubrahmaniyam2023} and of Methylene Blue at the TD-DFT/B97//3-21G level of theory in our MD simulations. The plots of the MSD$_w$'s are 
well fitted with Expression~\ref{eq:MSD_fit} (see Table~\ref{tab:regr_coef} for the values of a regression coefficient), allowing for an accurate estimation of the transport exponent $\beta$. As demonstrated in Figure~\ref{fig:transport_static}\textbf{c}-\textbf{d}, 
for both disorder strengths, the transport exponent, within the error, is close to two within the whole range of lower polariton states with a well-defined wave vector, implying that the transition between ballistic transport and diffusion is not due to static excitation energy disorder.

\begin{table}[htpb]
\centering
\caption{Coefficients of determination ($R^2$) for the fits to Expression~\ref{eq:MSD_fit} of the MSD$_w$'s obtained for the partial photonic wave function, $|\Psi^{\text{part}}_{\text{phot},w}(z,t)\rangle$, in 
static simulations of two-level systems with  disorder strength $\sigma$ of 22~meV (second column) and 63~meV (third column). The first column indicates ranges of wave vectors ($k_z$) at which partial wave functions were extracted. From the last two windows, the partial photonic wave function was extracted only in simulations with $\sigma=22$~meV, because in the case of $\sigma=63$~meV, the wave vectors of lower polaritons corresponding to these windows are not well-defined.}
\begin{footnotesize}
\begin{tabular}{c | c | c}
\hline
  $k_z$-range ($\mu$m$^{-1}$) & $R^2$ ($\sigma=22$~meV) & $R^2$ ($\sigma=63$~meV) \\
  \hline
  $\left[9.00; 9.50\right]$ & 1.0000 & 1.0000  \\
  $\left[9.25; 9.75\right]$ & 1.0000 & 0.9998 \\
  $\left[9.50; 10.00\right]$ & 1.0000 & 0.9986 \\
  $\left[9.75; 10.25\right]$ & 1.0000 & 0.9986 \\
  $\left[10.00; 10.50\right]$ & 1.0000 & 0.9983 \\
  $\left[10.25; 10.75\right]$ & 1.0000 & 0.9998 \\
  $\left[10.50; 11.00\right]$ & 1.0000 & 0.9989 \\
  $\left[10.75; 11.25\right]$ & 1.0000 & -- \\
  $\left[11.00; 11.50\right]$ & 1.0000 & -- \\
\hline
\end{tabular}
\end{footnotesize}
\label{tab:regr_coef}
\end{table}

\clearpage

\subsection{MD simulations with constraints on bond lengths and out-of-plane motions}
\label{sec:suppressed_DOF}

As we discuss in the main text, molecular vibrations are crucial for the observed crossover between the two regimes of polariton transport, because this crossover is caused by  population exchange between bright and dark states, that is driven by nonadiabatic coupling. To provide further support for this conclusion, we also performed simulations with a reduced number of degrees of freedom. Specifically, we computed trajectories of $N=1024$ Methylene Blue molecules in vacuum, with  constraints on all bond-lengths,\cite{Hess1997} as well as on the out-of-plane motions of the heavy (carbon, nitrogen, oxygen and sulfur) atoms.
Under these conditions, the excitation energy disorder strength was reduced from $\sigma=63$~meV to $\sigma=27$~meV.

\begin{figure}[!htb]
\centering
\includegraphics[width=0.93\textwidth]{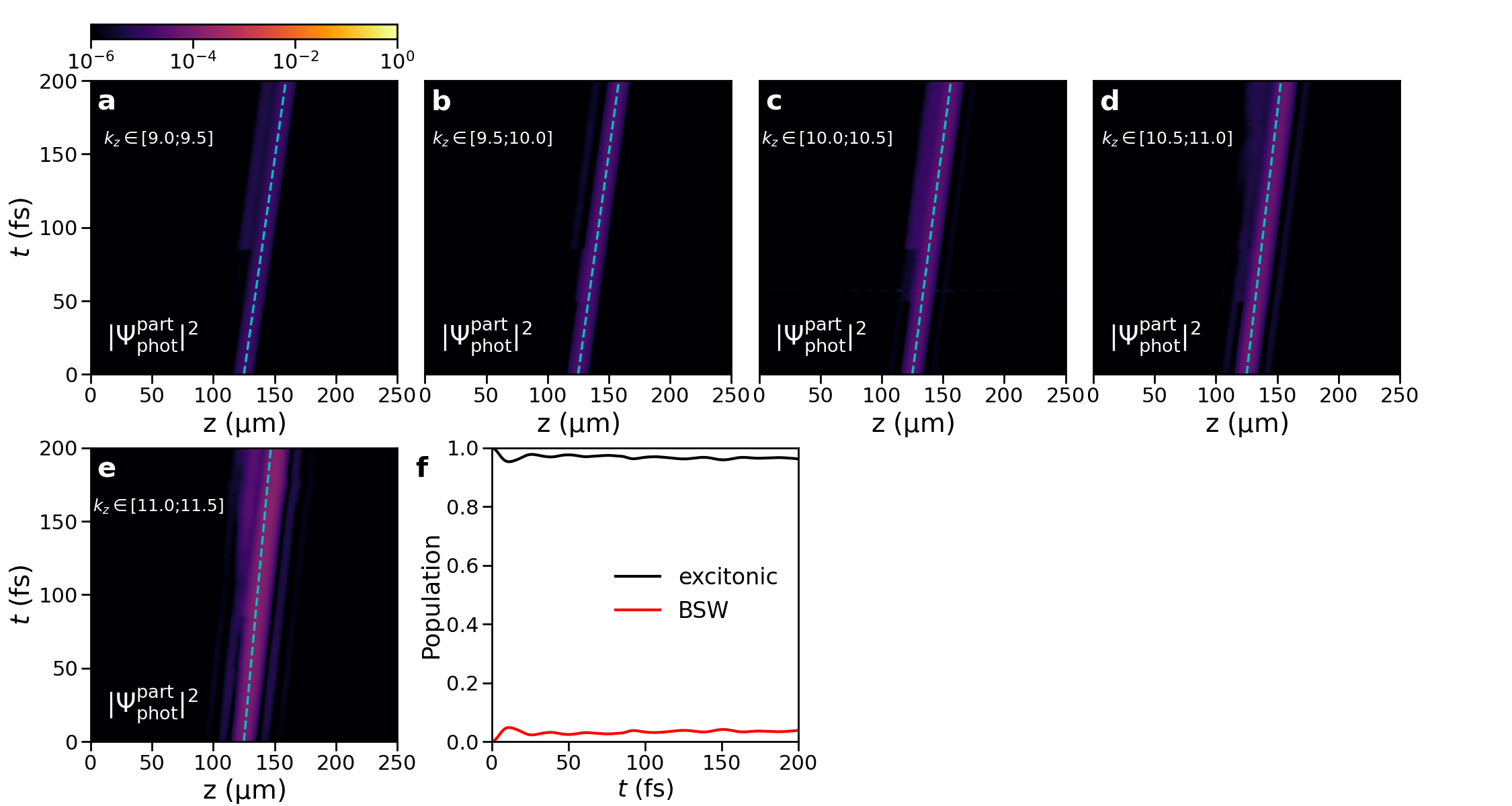}
  \caption{Panels~\textbf{a}-\textbf{g}: Time-space maps of the probability amplitude of the partial photonic wave function $|\Psi^{\text{part}}_\text{phot}(z,t)|^2$ extracted through different wave vector windows in simulations with constraints imposed on the nuclear degrees of freedom. Cyan lines correspond to the central group velocity of LP of each window. Panel~\textbf{h}: Contributions of molecular excitons (black) and BSW excitations (red) to $|\Psi(z,t)|^2$.}
  \label{fig:BSW_transport_rest}
\end{figure}

Figure~\ref{fig:BSW_transport_rest}\textbf{f} depicts the contributions of the MeB excitations and of the BSW modes to the total wave function $|\Psi(t)|^2$ as a function of time. Throughout the simulation, the excitonic contribution remains above 90\%, which implies that the imposed restrictions result in a much more modest population transfer between bright and dark states with the majority of the population remaining in the dark states, which are mostly composed of molecular excitons. The lower population transfer dynamics reflects a reduction of the nonadiabatic coupling\cite{Tichauer2022} via (i) a suppression of the molecular motions, and (ii) a more narrow energy distribution of the dark state manifold. As the non-adiabatic coupling depends on the overlap between the non-adiabatic coupling vector and the molecular displacements, reducing the latter decreases the population transfer. Likewise, since the non-adiabatic coupling vector for population transfer between dark and bright states is inversely proportional to the energy gap between these states~\cite{Yarkony2012}, narrowing the distribution of the dark states increases that gap and thus also decreases the non-adiabatic population transfer.


\begin{figure}[!htb]
\centering
\includegraphics[width=0.93\textwidth]{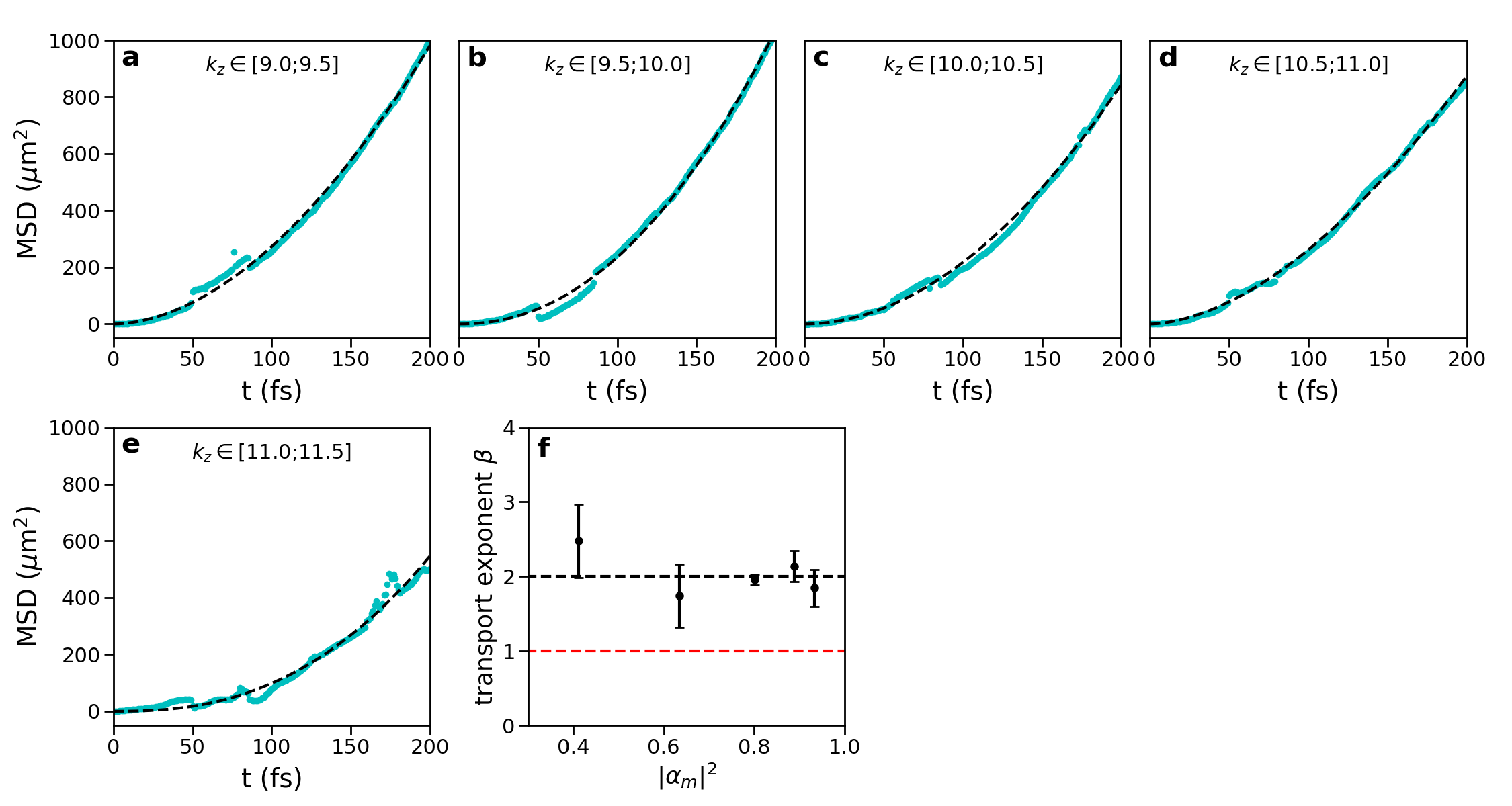}
  \caption{Panels~\textbf{a}-\textbf{g}: Mean squared displacement, $\text{MSD}_w(t)-\text{MSD}_w(0)$, of partial photonic wave functions $|\Psi^{\text{part}}_{\text{phot},w}|^2$ as a function of time, extracted from different wave vector windows in simulations with constraints on bond-lengths and out-of-plane motions. Dashed lines are fits to $D_{\beta}\cdot t^\beta$. Panel~\textbf{h}: The transport exponent $\beta$ as a function of the BSW modes contribution $|\alpha_{\text{m}}|^2$ to polaritonic states. The error bars are standard deviations of five runs.}
  \label{fig:MSD_part_phot_rest}
\end{figure}

Suppression of population exchange between bright and dark states leads to ballistic transport of all states in the LP branch. This is evident from the plots of the probability amplitudes of the partial photonic wave functions $|\Psi^{\text{part}}_{\text{phot},w}(z,t)|^2$ associated with the different wave vector windows in Figure~\ref{fig:BSW_transport_rest}\textbf{a}-\textbf{e}.
In all windows, the population propagates with the central group velocity of that window (cyan lines in Figure~\ref{fig:BSW_transport_rest}\textbf{a}-\textbf{e}). Accordingly, the MSD$_w$ plots in Figure~\ref{fig:MSD_part_phot_rest}\textbf{a}-\textbf{e} display a quadratic time-dependence for all wave vector windows. Fitting 
Equation~\ref{eq:MSD_fit} to these MSD$_w$'s reveals that the transport coefficients  are always close to two (Figure~\ref{fig:MSD_part_phot_rest}\textbf{f}), suggesting ballistic motion that is independent of the photonic content of the LP states.
We note that similar results were obtained if the partial wave function contains both excitonic and photonic parts (\textit{i.e.}, $|\Psi^{\text{part}}(z,t)|^2$, Equation~\ref{eq:WP_total_partial}).

\clearpage

\subsection{Simulation of two-level systems with quasi-dynamic disorder}

\begin{figure}[!htb]
\centering
\includegraphics[width=1\textwidth]{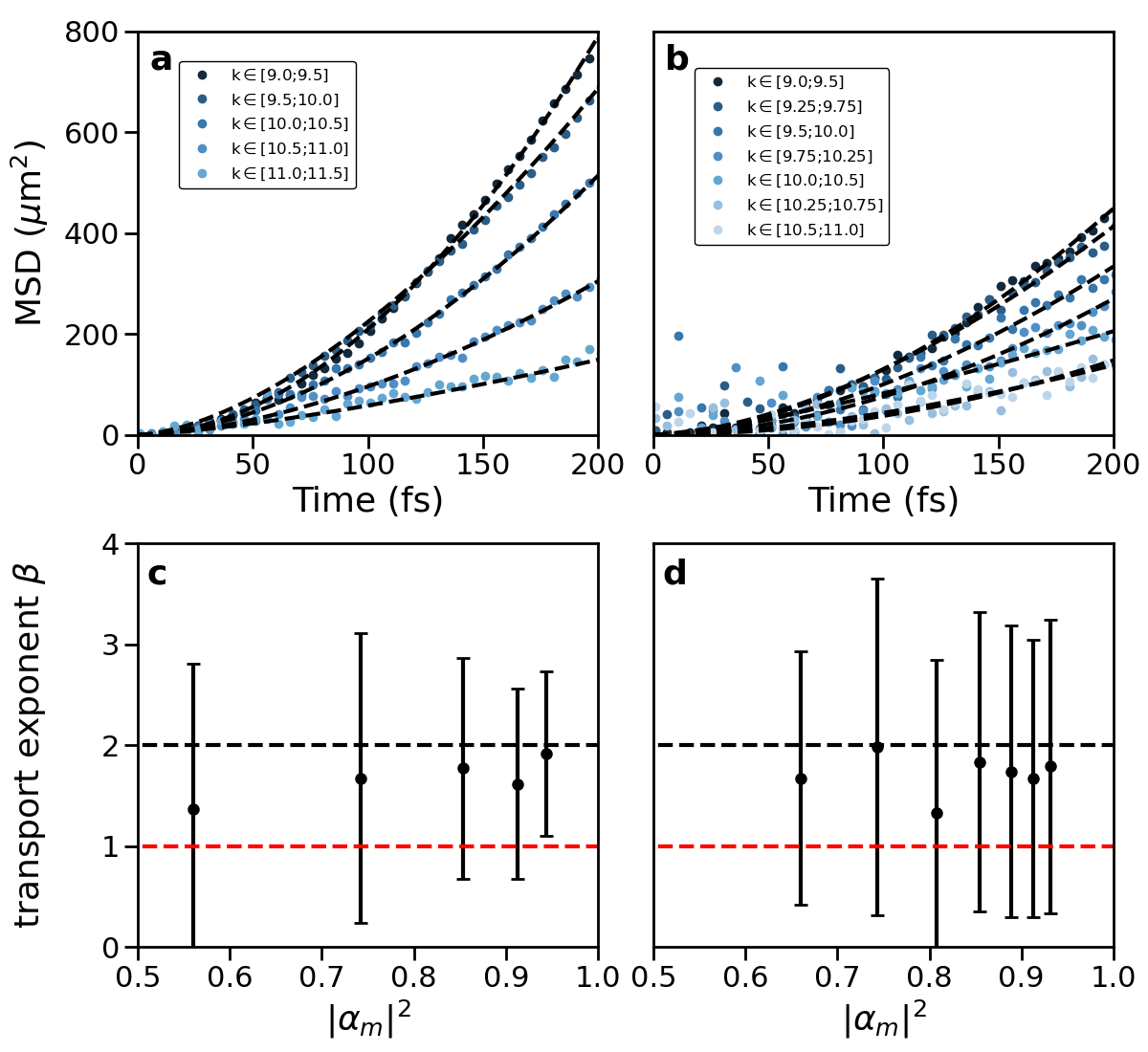}
  \caption{Panels~\textbf{a}~and~\textbf{b}: Mean squared displacement (MSD$_w$) of the partial photonic wave function $|\Psi^{\text{part}}_{\text{phot}}|^2$ extracted from different wave vector windows in quasi-dynamic simulations of two-level systems with excitation energy disorder of $\sigma=22$~meV (\textbf{a}) and $\sigma=63$~meV (\textbf{b}). Dashed lines are fits to $D_{\beta}\cdot t^\beta$ Panels~\textbf{c}~and~\textbf{d}: Values of the transport exponent, $\beta$, as a function of the BSW modes contribution $|\alpha_{\text{m}}|^2$ to polaritonic states. The errors are standard deviations of a hundred individual runs.}
  \label{fig:transport_quasidyn}
\end{figure}

To further confirm that the observed crossover in the transport regimes is not entirely due to structural disorder, we performed simulations of two-level systems with quasi-dynamic excitation energy disorder. In these simulations, excitation energies of the two-level systems were drawn from a Gaussian distribution (Equation~9 in the main text) and resampled every five femtoseconds of the simulation\cite{Nitzan2023}. In Figure~\ref{fig:transport_quasidyn}\textbf{a}-\textbf{b}, the $\text{MSD}_w$ of the partial photonic wave function $|\Psi^{\text{part}}_{\text{phot},w}|^2$ extracted from different wave vector windows in simulations with disorder strengths of $\sigma=22$~meV and $\sigma=63$~meV, is presented. By fitting these $\text{MSD}_w$'s with Expression~\ref{eq:MSD_fit}, we extracted the values of the transport exponent, which are plotted in Figure~\ref{fig:transport_quasidyn}\textbf{c}-\textbf{d} as a function of the photonic Hopfield coefficients. Unlike the atomistic MD simulations, the result of the quasi-dynamic simulations of two-level systems do not allow us to clearly distinguish a decreasing trend in the transport exponent when the excitonic content of the LP states increases. 
Moreover, although for most of the windows, the value of the transport exponent remains closer to two, thus suggesting ballistic transport, a large error of the values of $\beta$ does not allow to conclude with certainty which transport regime is operational in each of the wave vector windows. Nevertheless, because there is no clear crossover in the value of the transport exponent, these additional simulations further underline the role of vibrationally driven non-adiabatic population transfer in turning the propagation of polaritons with high excitonic content into a diffusion process.


\clearpage

\subsection{Simulation of $N=120$ molecules}
\label{sec:120mol}

\begin{figure}[!htb]
\centering
\includegraphics[width=1\textwidth]{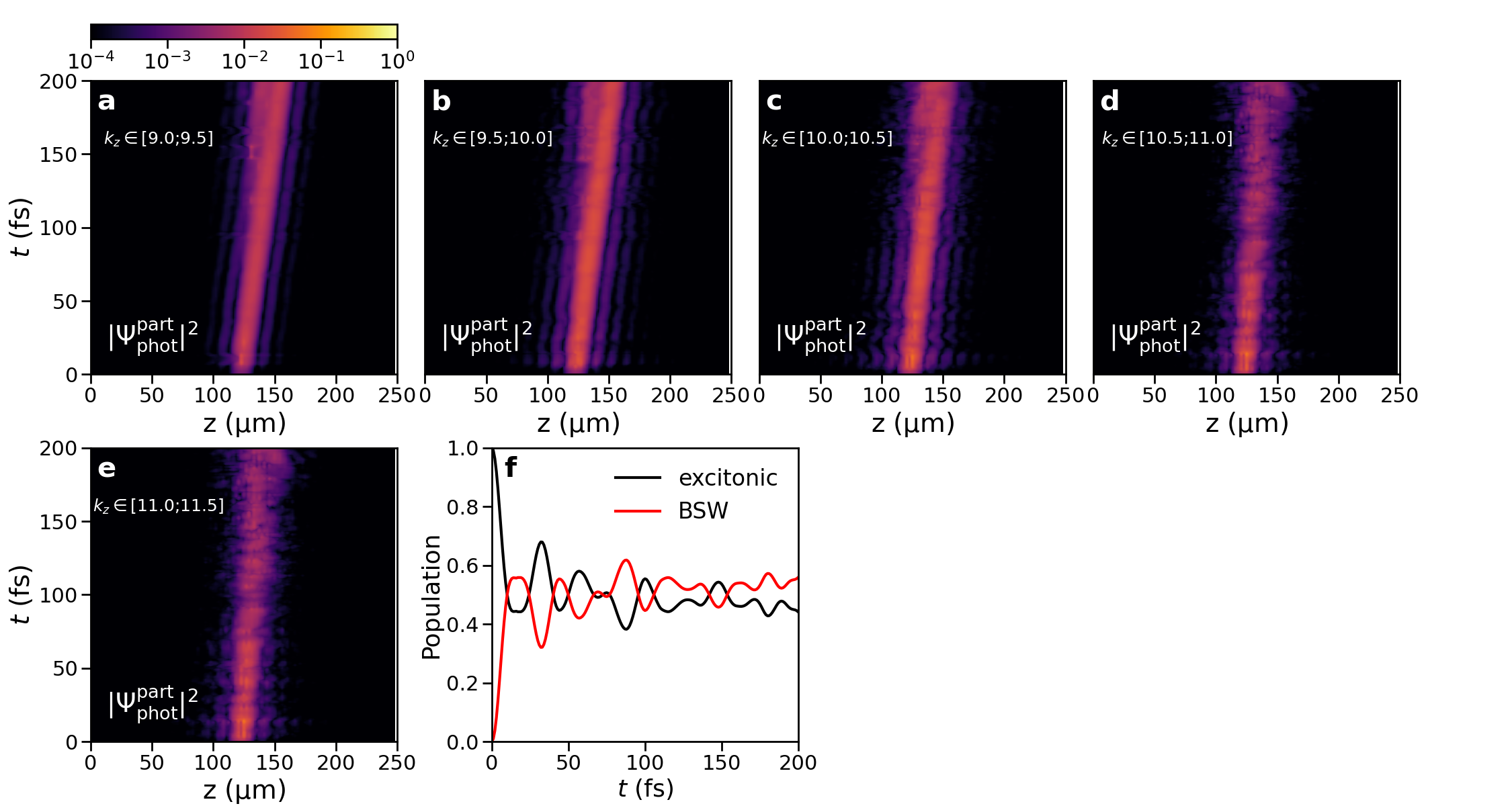}
  \caption{Panels~\textbf{a}-\textbf{g}: Time-space maps of the probability amplitude of the partial photonic wave function $|\Psi^{\text{part}}_{\text{phot},w}(z,t)|^2$ extracted from different wave vector windows, $w$, in simulations with $N=120$ molecules. Panel~\textbf{h}: Contributions of molecular excitons (black) and BSW excitations (red) to the total wave function ($|\Psi(z,t)|^2$).}\label{fig:BSW_transport_120mol}
\end{figure}

To understand how the density of molecular states influences the crossover between ballistic and diffusive propagation, we performed simulations with $N=120$ molecules coupled to $n_{\text{modes}}=120$ BSW modes. If this system were ideal and had no disorder, all $N+n_{\text{modes}}=240$ states would be polaritonic with each state having a both excitonic and photonic  contributions
(Figure~\ref{fig:dispersion}\textbf{a}). In realistic systems, however, there is always disorder, and the polaritonic states have energies that overlap with the distribution of molecular excitations. Due to that overlap, such states have an 
enhanced excitonic content and are hence  "grey"\cite{Houdre1996,Mony2021,Perez-Sanchez2024a,Dutta2023}. Nevertheless, for a system with as many molecules as BSW modes, the number of such grey states is limited, and one could expect a reduced contribution of molecular excitons to the total wave function, $|\Psi(t)|^2$, as compared to a system with $N>n_{\text{modes}}$. Indeed, Figure~\ref{fig:BSW_transport_120mol}\textbf{f} shows that the excitonic content of the total wave function in simulation with $N=120$ molecules remains around $50\%$ throughout the simulation, which is much smaller than the value of $\sim80\%$ in simulation with $N=1024$ molecules (Figure~\ref{fig:BSW_transport_phot}\textbf{h}).

\begin{figure}[!htb]
\centering
\includegraphics[width=1\textwidth]{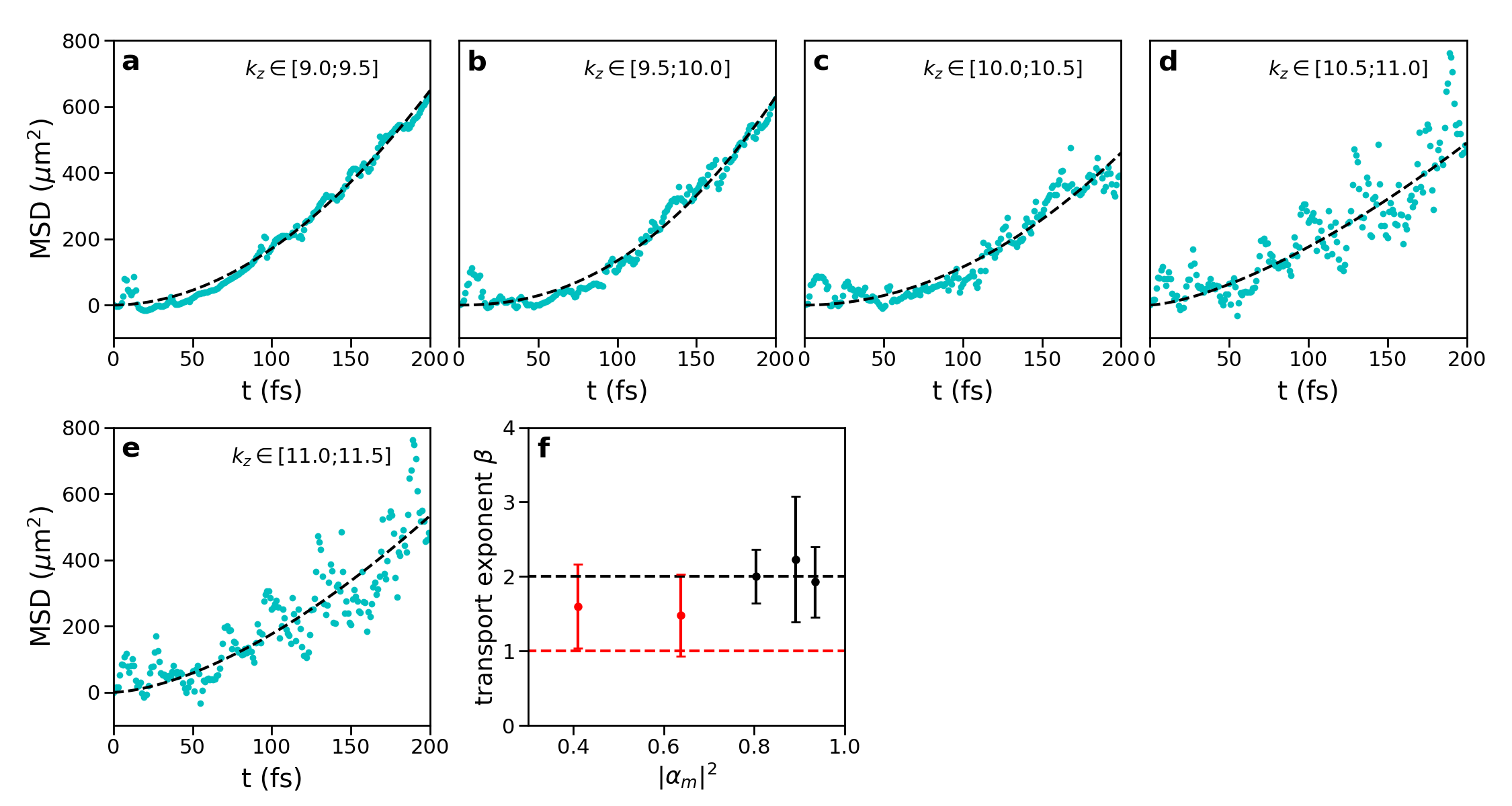}
  \caption{Panels~\textbf{a}-\textbf{g}: Mean squared displacement, $\text{MSD}_w(t)-\text{MSD}_w(0)$, of partial photonic wave functions $|\Psi^{\text{part}}_{\text{phot},w}|^2$ as a function of time, extracted from different wave vector windows in simulations with $N=120$ molecules. Dashed lines are fits of $D_{\beta}\cdot t^\beta$ to the cyan dots. Panel~\textbf{h}: The transport exponent $\beta$ as a function of the BSW modes contribution $|\alpha_{\text{m}}|^2$ to polaritonic states. The error bars are standard deviations of five runs.}
  \label{fig:MSD_part_phot_120mol}
\end{figure}

In 
Figures~\ref{fig:BSW_transport_120mol}\textbf{a}-\textbf{e}~and~\ref{fig:MSD_part_phot_120mol}\textbf{a}-\textbf{e}, we show the time-space maps of the probability amplitude of the partial photonic wave function, $|\Psi^{\text{part},w}_\text{phot}(z,t)|^2$, and the corresponding MSD$_w$'s, respectively, extracted from the same windows as in Section~\ref{sec:suppressed_DOF}. Fitting Expression~\ref{eq:MSD_fit} to these MSD$_w$'s yields the transport exponent for each window. These transport coefficients are plotted in Figure~\ref{fig:MSD_part_phot_120mol}\textbf{f} as a function of photonic content. The values of $\beta$ suggest that the transition to the diffusion regime remains incomplete, 
as even for states with the lowest photonic fractions, the transport exponent is in between one and two, indicating an intermediate regime\cite{Balasubrahmaniyam2023}. This finding underscores the role of the population transfers between bright and dark states, as the crossover between ballistic transport and diffusion can only be fully captured in a system with sufficiently many dark states, as was the case in simulations with $N=1024$ molecules (Figure~\ref{fig:MSD_part_phot}). 

\bibliography{main}